\documentclass[12pt]{article}

\usepackage{amsmath,natbib,mathrsfs}
\usepackage{amssymb,color}
\usepackage{appendix}
\usepackage{amsthm}
\usepackage{enumerate}
\usepackage{hyperref}
\usepackage{mathtools} 
\mathtoolsset{showonlyrefs=true}  
\usepackage{bm}
\usepackage{setspace}
\usepackage[margin=0.9in]{geometry}
 
 \newcounter{cases}
 \newenvironment{mycase}
 {
 	\setcounter{cases}{0}
 	\newcommand{\case}
 	{
 		\par\noindent\stepcounter{cases}\textbf{Case \thecases.}
 	}
 }
 {
 	\par
 }
 \renewcommand*\thecases{\arabic{cases}}
 
 \DeclareMathOperator*{\argmax}{arg\,max}
 \DeclareMathOperator*{\argmin}{arg\,min}

\numberwithin{equation}{section}
 
\theoremstyle{plain}
 
\newtheorem{theorem}{Theorem}[section]
\newtheorem{cor}[theorem]{Corollary}

\newtheorem{proposition}{Proposition}[section]

\theoremstyle{definition}
\newtheorem{remark}{Remark}[section]

 \newcommand{\upperRomannumeral}[1]{\uppercase\expandafter{\romannumeral#1}}

\newcommand{\mbp}{\mathbb{P}}
\newcommand{\mbq}{\mathbb{Q}}
\newcommand{\mbr}{\mathbb{R}}
\newcommand{\EP}{\mathbb{E}^{\mathbb{P}}}

\begin{document}
\author{Tim Leung\thanks{Department of Applied Mathematics, Computational Finance \& Risk Management Program, University of Washington, Seattle WA 98195.
\mbox{timleung@uw.edu}.},  \and Hyungbin Park\thanks{Department of Mathematical Sciences and Research Institute of Mathematics, Seoul National University, 1 Gwanak-ro, Gwanak-gu, Seoul 08826, Republic of Korea. \mbox{hyungbin@snu.ac.kr, hyungbin2015@gmail.com}}, \and Heejun Yeo\thanks{Department of Mathematical Sciences, Seoul National University, 1 Gwanak-ro, Gwanak-gu, Seoul 08826, Republic of Korea. \mbox{yeohj@snu.ac.kr}}}

\title{Robust Long-Term Growth Rate of Expected Utility for Leveraged ETFs}
\date{\today}   

\maketitle

\begin{abstract}
This paper analyzes the robust long-term growth rate of expected utility and expected return
from holding a leveraged exchange-traded fund (LETF). When the Markovian model parameters in the reference asset are uncertain, the robust long-term growth rate is derived by analyzing the worst-case parameters among an uncertainty set. We compute the growth rate and describe the optimal leverage ratio maximizing the robust long-term growth rate. To achieve this, the worst-case parameters are analyzed by the comparison principle, and the growth rate of the worst-case is computed using the martingale extraction method. The robust long-term growth rates are obtained explicitly under a number of models for the reference asset, including the geometric Brownian motion (GBM),
Cox--Ingersoll--Ross (CIR), 3/2, and Heston and 3/2 stochastic volatility models. Additionally, we demonstrate the impact of stochastic interest rates, such as the Vasicek and inverse GARCH
short rate models.
This paper is an extended work of \citet{Leung2017}.
\end{abstract}

\newpage
\section{Introduction} 
\subsection{Overview}
In recent years, leveraged exchange-traded funds (LETFs) have garnered popularity as a relatively new financial product. Unlike mutual funds, ETFs are specifically designed to track an index as closely as possible. Therefore,  the main purpose of an ETF is to give investors a pre-specified exposure to an index. For example, the world's largest ETF is the Standard \& Poor's Depositary Receipts (SPDR) S\&P 500 ETF (ticker: SPY), which seeks to replicate the daily performance of the S\&P 500 index.  Leveraged ETFs are designed to replicate a constant multiple, called \textit{leverage ratio},  of the daily returns of a reference index. The most common leverage ratios are $\{-3, -2, -1, 2, 3\}$. Hence, LETFs offer investors easy access to trade an index with different leveraged exposures, thereby amplifying returns and risks.

Mathematically and empirically, it has been shown that LETFs tend to suffer from value erosion over time, called volatility decay, which is proportional to the realized variance of the reference index, depending on the leverage ratio. Therefore, the question of what the long-term growth rate of an LETF is becomes complicated. This issue is particularly challenging when the parameters of the reference asset are uncertain.


In this study, we investigate the robust long-term growth rate of expected utility and expected return from holding an LETF under uncertain parameters of a reference asset.
To determine the robust long-term growth rate, we find the worst-case parameters out of an uncertainty set using the comparison principle and the martingale extraction method. Furthermore, we describe an optimal leverage ratio that maximizes the robust long-term growth rate. The reference asset considered for our analysis comprises a variety of models, including the GBM, CIR, 3/2 models, and Heston and 3/2 stochastic volatility models. Additionally, we explore the impact of stochastic interest rate models such as the Vasicek and inverse GARCH short rate model.

This study is a novel extension of the work conducted by \citet{Leung2017}, which presents explicit representations for the robust long-term growth rates under various models of the reference asset. The results of the current study have practical implications for investors who hold LETFs and seek to minimize the risks associated with uncertain parameters of the reference asset. For LETF selection and other purposes, our analysis also determines the optimal leverage ratio maximizing the robust long-term growth rate.
 
\subsection{Related literature}
Since the introduction of LETFs in 2006, their market has grown rapidly, promoting active research and development. For the price dynamics of general LETFs, \citet{chengLETF}, \citet{AZLETF}, and \citet{jarrow2010understanding} examined how the leverage ratio and realized variance of the underlying reference index can proportionally contribute to the erosion of an LETF's returns. \citet{leung2014commodity} analyzed the tracking performance of LETFs on a wide array of commoditiesby specifically focusing on the effects of volatility decay as LETFs are known to be highly susceptible to this phenomenon. \citet{lee2015how} and \citet{LeungLorig2016} studied the relationship between ETF and LETF implied volatility surfaces. Furthermore, the study of options written on LETF has appealed several researchers, including \citet{LeungSircar2015}, \citet{trainor2016leveraged}, and \citet{lopez2018short}.

The robust framework discussed in this paper can be considered as an extension of the uncertain volatility model, which assumes volatility as an unknown but it lies on a given finite interval, where it imposes uncertainty on as many parameters as possible. \citet{avellaneda1995pricing} proposed a model for pricing and hedging derivative securities and option portfolios under volatility uncertainty, and described the prices of derivative assets by a nonlinear PDE, called the Black--Scholes--Barenblatt equation. \citet{matoussi2015robust} studied the problem of robust utility maximization in an incomplete market with volatility uncertainty using second-order backward stochastic differential equations (2BSDEs) with quadratic growth generators. They demonstrated that the value function can be expressed as the initial value of a particular 2BSDE for exponential, power, and logarithmic utilities. Furthermore, robust utility maximization has been studied in more general settings. \citet{tevzadze2013robust} set both volatility and trend as uncertain, and derived explicitly charaterized solutions through the Hamilton--Jacobi--Bellman--Isaacs equation. \citet{neufeld2018robust} presented a semi-closed form solution for an optimal investment strategy and a worst-case model analysis for an investor with logarithmic or power utility under model uncertainty specified by a set of possible Lévy triplets. For robust methods applied to LETF, \citet{cox2019robust} conducted research on robust hedging of LETF options.

\section{LETF price dynamics}
\label{sec:LETF_dynamics}

Let $(\Omega,\mathcal{F},\mathbb{F},\mathbb{P})$ be a filtered probability space where the filtration $\mathbb{F}:=(\mathcal{F}_t)_{t\ge 0}$ is generated by a
$d$-dimensional standard Brownian motion $B$. Assume that a reference index $X$ is a diffusion process given by an equation
\begin{align}\label{dXt1}
	\frac{dX_t}{X_t}=\mu_t\,dt+\sigma_t\, dB_t\,,  \qquad t\ge 0\,,
\end{align}
where both the drift $\mu=(\mu_t)_{t\ge0}$ and the diffusion process $\sigma=(\sigma_t^1,\cdots,\sigma_t^{d})_{t\ge0}$ are $\mathbb{F}$-adapted. An LETF is a portfolio designed to maintain a constant proportional exposure to the reference $X$. The constant proportion
is called a leverage ratio and is denoted by $\beta$. 
In various countries, the leverage ratio of an ETF is influenced by leverage regulation policies. Thus, we set the leverage ratio range from $\underline{\beta}$ to $\overline{\beta},$ where $\underline{\beta}<0$ and $\overline{\beta}>1$. Let $L_t$ be the LETF price at time $t\geq 0$. At time $t$, an agent invests an amount  $\beta L_t$ in $X$ (short position of the amount $|\beta|L_t$ if $\beta<0$) and
$(1-\beta)L_t$ in a bank account with short rate $r.$ Then 
\begin{equation*}
	\begin{aligned}
		\frac{dL_t}{L_t}
		&=\beta\left(\frac{dX_t}{X_t}\right)-((\beta-1)r_t)\,dt\\
		&=(\beta \mu_t-(\beta-1)r_t)\,dt+\beta\sigma_t\, dB_t\;.
	\end{aligned}
\end{equation*}
Without loss of generality, we set $L_0=X_0=1.$ 

The expected utility from holding the LETF up to time $T$ is given by 
\begin{equation}
	\begin{aligned}\label{EL3} 
		\mathbb{E}^\mathbb{P}[L_T^p ]&=\mathbb{E}^\mathbb{P}[X_T^{p \beta} e^{\int_0^T(-p (\beta-1)r_s-\frac{1}{2}p \beta(\beta-1)|\sigma_s|^2)\,ds}] \\
		&=\mathbb{E}^\mathbb{P}[e^{{\int_0^T(p\beta\mu_s-p(\beta-1)r_s-\frac{1}{2}p\beta^2|\sigma_s|^2)\,ds+p\beta\int_0^T\sigma_s\,dB_s}}]\,.
	\end{aligned}
\end{equation}
Supposing a local martingale $\mathcal{E}\bigl(p\beta\int_0^{\cdot}\sigma_s\,dB_s\bigl)$ as a martingale, we can define a new measure $\hat{\mathbb{P}}_t$ on $\mathcal{F}_t$ by  
\begin{align}
	\frac{d\hat{\mathbb{P}}_t}{d\mathbb{P}}=\mathcal{E}\left(p\beta\int_0^{\cdot}\sigma_s\,dB_s\right)_t,\label{dphatdp}
\end{align}
for each $t\geq 0$. By the Girsanov theorem, the process  $\hat{B}$ defined as
\begin{equation}\label{eqn:BM_under_hat_P}
	\hat{B}_s:=-p\beta\int_0^s\sigma_u\,du+B_s\quad,0\leq s\leq t
\end{equation} 
is a standard Brownian motion under $\mathbb{\hat{\mathbb{P}}}_t.$ Note that $(\hat{\mathbb{P}}_{s})_{s\geq 0}$ is consistent in the sense that $\hat{\mathbb{P}}_{T}|_{\mathcal{F}_t}=\hat{\mathbb{P}}_{t}$ for all $T\geq t\geq 0$. Thus, writing $\hat{\mathbb{P}}$ without the index $t$ would not cause any confusion. Similarly, \eqref{eqn:BM_under_hat_P} can be regarded as a standard Brownian motion under $\hat{\mathbb{P}}$ on any finite time horizon. Utilizing the universal notations of $\hat{\mathbb{P}}$ and $\hat{B}$ and applying \eqref{eqn:BM_under_hat_P} to \eqref{dXt1} and \eqref{EL3}, we have
$$\frac{dX_t}{X_t}=(\mu_t+p\beta|\sigma_t|^2)\,dt+\sigma_t\,d\hat{B}_t,\quad 0\leq t\leq T,$$
and 
\begin{equation}\label{eqn:expected_utility_of_LETF_hat_P}
	\begin{aligned}
		&\mathbb{E}^\mathbb{P}[L_T^p]
		=\mathbb{E}^\mathbb{\hat{\mathbb{P}}}\left[e^{\int_0^T(p\beta\mu_s -p(\beta-1) r_s-\frac{1}{2}p(1-p)\beta^2 |\sigma_s|^2)\,ds}\right]\;,
	\end{aligned}
\end{equation}
for any $T>0$.

\section{Uncertainties on reference process} \label{sec:Uncertain_ref}
Throughout this section, we assume the interest rate as constant $r$ and initial value $X_0$ of the reference process as $1.$ Let $\alpha=(\alpha_1,\ldots,\alpha_n)\in\mbr^n$ be the set of model parameters. That is, the drift and diffusion terms of $X$ depend on $\alpha$. We write dynamics of \textit{X} as
\begin{equation}
	\frac{dX_t^{\alpha}}{X_t^{\alpha}}=\mu(X_t^{\alpha};\alpha)dt+\sigma(X_t^{\alpha};\alpha)\,dB_t,\quad X_0=1,
\end{equation}
to point out dependence of $\mu$ and $\sigma$ (and thus $X$) on $\alpha$.
Correspondingly, the expected utility in \eqref{EL3} can be written as
\begin{equation}
	\EP[(L_t^{\alpha})^p]=e^{-p\,r (\beta-1)t}\mathbb{E}^\mathbb{P}[(X_t^\alpha)^{p \beta} e^{-\frac{1}{2}p\beta(\beta-1)\int_0^t\sigma(X_s^{\alpha};\alpha)^2\,ds}],\quad t\geq 0.
\end{equation}

Assuming $\alpha$ as uncertain, but knowing that  $\alpha$  ranges over a compact rectangle in $\mathbb{R}^d$, there exist two vectors $\underline{\alpha}=(\underline{\alpha}_1,\cdots,\underline{\alpha}_n)\in\mbr^n$ and $\overline{\alpha}=(\overline{\alpha}_1,\cdots,\overline{\alpha}_n)\in\mbr^n$ such that $\underline{\alpha}_i\leq\overline{\alpha}_i$ for all $i=1,\ldots,n$ and
\begin{equation}
	\alpha\in[\underline{\alpha},\overline{\alpha}]:=\prod_{i=1}^{n}[\underline{\alpha}_i,\overline{\alpha}_i]\,.
\end{equation}
We investigate the growth rate of the worst-case expected utility
\begin{equation} \label{growth_rate:worst}
	\frac{1}{T}\log{\inf_{\alpha\in[\underline{\alpha},\overline{\alpha}]}\EP[(L_T^{\alpha})^p]}
\end{equation}
as $T\to\infty$  and find the value of the limit when it converges. 
The dependence of the process $X$ on $\alpha$ is denoted in the expectation  $\mathbb{E}^{\mathbb{P}^\alpha}[\cdot]$, thereby we can simply write $X_t = X^\alpha_t$. For example,
\begin{equation}
	\mathbb{E}^{\mbp^{\alpha}}[X_T^{p \beta} e^{-\frac{1}{2}p\beta(\beta-1)\int_0^T\sigma^2(X_s)\,ds}]=\mathbb{E}^\mathbb{P}[(X_t^\alpha)^{p\beta} e^{-\frac{1}{2}p\beta(\beta-1)\int_0^t\sigma^2(X_s^{\alpha};\alpha)\,ds}] \,.
\end{equation}

Next, we define the worst-case expectation as
\begin{align}
	v_T&:=\inf_{\alpha\in[\underline{\alpha},\overline{\alpha}]}\mathbb{E}^{\mbp^{\alpha}}[L_T^p]=e^{-p\,r(\beta-1)T}\inf_{\alpha\in[\underline{\alpha},\overline{\alpha}]}V(T\,;\,\alpha), \label{eqn:V,v:reference}
\end{align}
with \[V(T\,;\,\alpha):=\mathbb{E}^{\mbp^{\alpha}}[X_T^{p \beta} e^{-\frac{1}{2}p\beta(\beta-1)\int_0^T\sigma^2(X_s)\,ds}].\]
Then \eqref{growth_rate:worst} can be written to have the expression
\begin{equation}
	\lim_{T\to\infty}\frac{1}{T}\log{v_T}=-p\,r(\beta-1)+\lim_{T\to\infty}\frac{1}{T}\log{\inf_{\alpha\in[\underline{\alpha},\overline{\alpha}]}V(T\,;\,\alpha)}.
\end{equation} \vspace{2mm}

A strategy for analyzing the worst-case expected utility includes the following.
The comparison principle for SDEs can be used to find $\alpha^*\in[\underline{\alpha},\overline{\alpha}]$ and a constant $C>0$ such that
\begin{equation}
	v_T\geq Ce^{-p\,r(\beta-1)T}V(T\,;\,\alpha^*).
\end{equation}
Then, the inequalities achieved include
\begin{align}
	-p\,r(\beta-1)+\limsup_{T\to\infty}\frac{1}{T}\log{V(T\,;\,\alpha^*)}
	&\geq \limsup_{T\to\infty}\frac{1}{T}\log{v_T} \\
	&\geq \liminf_{T\to\infty}\frac{1}{T}\log{v_T} \\
	&\geq -p\,r(\beta-1)+\liminf_{T\to\infty}\frac{1}{T}\log{V(T\,;\,\alpha^*)}.
\end{align}
Successively, the problem reduces to establish the equality
\begin{equation}
	\limsup_{T\to\infty}\frac{1}{T}\log{V(T\,;\,\alpha^*)}=\liminf_{T\to\infty}\frac{1}{T}\log{V(T\,;\,\alpha^*)}
\end{equation}
and find the value of the limit can be found as
\begin{equation}
	\lim_{T\to\infty}\frac{1}{T}\log{V(T\,;\,\alpha^*)}.
\end{equation}
Note that the worst-case set of parameters may vary depending on the leverage ratio $\beta$. As seen in \citet{Leung2017}, the martingale extraction method can be a suitable tool for this problem.

\subsection{GBM model}
For the first example, we consider the problems in the GBM model, where $X$ follows the SDE, such that
$$dX_t=\mu X_t\,dt+\sigma X_t\,dB_t\,,\;t\geq0$$
with $(\mu,\sigma)\in[\underline{\mu},\overline{\mu}]\times[\underline{\sigma},\overline{\sigma}],\,\underline{\mu},\underline{\sigma}>0$. 
For each pair of $(\mu,\sigma)$, the expected utility is given explicitly by
\begin{align} 
	\mathbb{E}^{\mbp^{\mu,\sigma}}[L_T^p]&=\mathbb{E}^\mathbb{P}[e^{p(\beta\mu-(\beta-1)r)T-\frac{1}{2}p\beta^2\sigma^2T+p\beta\sigma B_T}]\\
	&=\mathbb{E}^{\mathbb{P}}[e^{p\beta\sigma B_T-\frac{1}{2}p^2\beta^2\sigma^2 T}e^{p(\beta\mu-(\beta-1)r)T-\frac{1}{2}p(1-p)\beta^2\sigma^2T}]\\
	&=\mathbb{E}^{\mathbb{Q}}[e^{p(\beta\mu-(\beta-1)r)T-\frac{1}{2}p(1-p)\beta^2\sigma^2T}]\label{change_measure:GBM}  \\ 
	&=e^{p(\beta\mu-(\beta-1)r)T-\frac{1}{2}p(1-p)\beta^2\sigma^2T}, 
\end{align} 
where the probability measure $\mbq$ is defined on $\mathcal{F}_T$ by
\begin{equation}
	\frac{d\mathbb{Q}}{d\mathbb{P}}\biggl|_{\mathcal{F}_T}=e^{p\beta\sigma B_T-\frac{1}{2}p^2\beta^2\sigma^2 T}.
\end{equation}
Then,
\begin{equation} \label{comp:GBM}
	\inf_{(\mu,\sigma)\in[\underline{\mu},\overline{\mu}]\times[\underline{\sigma},\overline{\sigma}]}\mathbb{E}^{\mbp^{\mu,\sigma}}[L_T^p]\geq e^{p(\beta\mu^*(\beta)-(\beta-1)r)T-\frac{1}{2}p(1-p)\overline{\sigma}^2\beta^2T},
\end{equation}
where
\begin{equation}
	\mu^{*}(\beta)=\begin{cases}
		\underline{\mu}, &  \beta\geq 0 \\ \overline{\mu}, & \beta<0.
	\end{cases}
\end{equation}
Conversely, by the definition of infimum, we have
\begin{equation}
	\inf_{(\mu,\sigma)\in[\underline{\mu},\overline{\mu}]\times[\underline{\sigma},\overline{\sigma}]}\mathbb{E}^{\mbp^{\mu,\sigma}}[L_T^p]\leq\mathbb{E}^{\mbp^{\mu^*(\beta),\overline{\sigma}}}[L_T^p]=e^{p(\beta\mu^*(\beta)-(\beta-1)r)T-\frac{1}{2}p(1-p)\overline{\sigma}^2\beta^2T}.
\end{equation}
In conjunction with \eqref{comp:GBM}, we deduce that
\begin{equation} \label{growth_rate:GBM}
	\lim_{T\to\infty}\frac{1}{T}\log{\inf_{(\mu,\sigma)\in[\underline{\mu},\overline{\mu}]\times[\underline{\sigma},\overline{\sigma}]}\mathbb{E}^{\mbp^{\mu,\sigma}}}[L_T^p]=p\,r+p(\mu^*(\beta)-r)\beta-\frac{1}{2}p(1-p)\overline{\sigma}^2\beta^2. 
\end{equation}

Next, we determine the optimal leverage ratio $\beta^*$, for which we define a function $\Lambda$ of $\beta$ as
$$\Lambda(\beta):=p\,r+p(\mu^*(\beta)-r)\beta-\frac{1}{2}p(1-p)\overline{\sigma}^2\beta^2.$$
The optimal leverage ratio depends on the relationship between $r$ and uncertainty set $[\underline{\mu},\overline{\mu}]$; hence, we classify the relationship into three cases and find $\beta^*$ for each case. Moreover, distinction between the two cases, $\beta\geq 0$ and $\beta<0$, is necessary. Nevertheless, the calculation for each case is simple, as $\Lambda$ is a quadratic function with respect to $\beta$. \vspace{3mm}

\begin{mycase}
	\case $\overline{\mu}<r$:	If $\beta\geq 0$, then $\beta^*=0$. Otherwise, $\beta^*=\dfrac{\overline{\mu}-r}{(1-p)\overline{\sigma}^2}$, because
	\begin{equation}
		\Lambda(0)=p\,r<p\,r+\frac{1}{2}\frac{(\overline{\mu}-r)^2}{(1-p)\overline{\sigma}^2}=\Lambda\Big(\frac{\overline{\mu}-r}{(1-p)\overline{\sigma}^2}\Big).
	\end{equation}\vspace{3mm}
	
	\case $\underline{\mu}\leq r\leq\overline{\mu}$:
	Clearly, $\beta^*=0$. \vspace{3mm}
	
	\case $r<\underline{\mu}$:
	If $\beta\geq 0$, then $\beta^*=\dfrac{\underline{\mu}-r}{(1-p)\overline{\sigma}^2}$. $\Lambda$ does not attain a maximum on $[-5,0)$, and $\underset{\beta\in[-5,0)}{\sup}\Lambda(\beta)=\Lambda(0)=p\,r$. Thus, $\beta^*=\dfrac{\underline{\mu}-r}{(1-p)\overline{\sigma}^2}$. \vspace{3mm}
	
\end{mycase}

The results obtained are consistent with our intuition that an agent who considers the worst-case scenario will not invest in LETF unless the expected rate of return of the reference goes either higher or lower than the interest rate $r$ for every scenario. Clearly, a long (respectively, short) position in LETF can be taken if the worst (respectively, best) expected rate of return of the reference exceeds (respectively, is inadequate) $r$.

\begin{remark}
	The results remains valid even when $\mu$ and $\sigma$ are extended to progressively measurable processes. Indeed,   $(\sigma_t)_{t\geq 0}$ is bounded and the SDE 
	$$dX_t=\mu_tX_t\,dt+\sigma_tX_t\,dB_t\,,\;t\geq0$$
	has a unique strong solution. Thus, \eqref{change_measure:GBM} and \eqref{comp:GBM} stay applicable for extended $\mu$ and $\sigma$ with minor adjustments. Therefore, \eqref{growth_rate:GBM} holds for generalized $\mu$ and $\sigma$.
\end{remark}

\subsection{CIR model}
\label{sec:CIR}
As an interest rate model, Cox and Ross \citet{cox1985theory} proposed the well-known CIR model
\begin{equation}\label{SDE:CIR}
	dX_t=(b-a X_t)\,dt+\sigma\sqrt{X_t}\,dB_t  \,,
\end{equation}
with parameters $a,\sigma>0$ and $2b>\sigma^2$. In this model, we set $\underline{\alpha}=(\underline{b},\underline{a},\underline{\sigma}),\,\overline{\alpha}=(\overline{b},\overline{a},\overline{\sigma}),\,[\underline{\alpha},\overline{\alpha}]=[\underline{b},\overline{b}]\times[\underline{a},\overline{a}]\times[\underline{\sigma},\overline{\sigma}]$ with $\underline{a},\underline{\sigma}>0,\, \underline{b}>\overline{\sigma}^2$.
Then, the worst-case expected utility $v_T$ is given by
\begin{align}
	v_T&=e^{-p\,r(\beta-1)T}\inf_{\alpha\in[\underline{\alpha},\overline{\alpha}]}V(T\,;\,\alpha) \\
	&=e^{-p\,r(\beta-1)T}\inf_{\alpha\in[\underline{\alpha},\overline{\alpha}]}\mathbb{E}^{\mbp^{\alpha}}\bigl[ X^{p\beta}e^{-\frac{1}{2}p\beta(\beta-1)\int_0^T\frac{\sigma^2}{X_s}\,ds}\bigr].
\end{align}

Set $Y=X/\sigma^2$. Then, $Y$ is a solution to the SDE
\begin{equation}
	dY_t=\left(\frac{b}{\sigma^2}-aY_t\right)dt+\sqrt{Y_t}\,dB_t,\quad Y_0=1/\sigma^2,
\end{equation}
and $V$ satisfies
\begin{equation}  \label{V:CIR}
	V(T\,;\,\alpha)=\sigma^{2p\beta}\,\mathbb{E}^{\mbp^{\alpha}}\bigl[ Y_T^{p\beta}e^{-\frac{1}{2}p\beta(\beta-1)\int_0^T\frac{1}{Y_s}\,ds}\bigr].
\end{equation}
The eigenpair problem for the infinitesimal generator of $Y$ is given by 
\begin{equation}
	-\lambda\phi(y)=\frac{1}{2}y\phi''(y)+\left(\frac{b}{\sigma^2}-ay\right)\phi'(y)-\frac{1}{2}p\beta(\beta-1)\frac{1}{y}\phi(y).
\end{equation}
Then, a pair
\begin{equation} \label{eigpair:CIR}
	(\lambda(b,a,\sigma),\phi(y))=(a\,\eta(b,\sigma),\,y^{\eta(b,\sigma)}),
\end{equation}
where
\begin{equation} \label{eta:CIR}
	\eta(b,\sigma)=-\Big(\frac{b}{\sigma^2}-\frac{1}{2}\Big)+\sqrt{\Big(\frac{b}{\sigma^2}-\frac{1}{2}\Big)^2+p\beta(\beta-1)}\,,
\end{equation}
can be shown as a solution to the eigenpair problem. Here, $\eta$ is real because $\underline{b}>\overline{\sigma}^2$ is assumed. Since a local martingale 
\begin{equation} \label{process:loc_mart:CIR}
	M_t=\left(\frac{Y_t}{y}\right)^{\eta(b,\sigma)}\exp\left\{\lambda(b,a,\sigma) t-\frac{1}{2}p\beta(\beta-1)\int_0^t\frac{1}{Y_s}\,ds\right\},\quad 0\leq t\leq T
\end{equation}
is a true martingale \cite[Theorem 4.8.5 (ii)]{Pinsky1995}, we can define a probability measure $\mathbb{Q}^{b,a,\sigma}$ on $\mathcal{F}_T$ via
\begin{equation}
	\frac{d\mathbb{Q}^{b,a,\sigma}}{d\mathbb{P}}\biggl|_{\mathcal{F}_T}=M_T,
\end{equation}
under which the process $Y$ satisfies $$dY_t=\left(\frac{b}{\sigma^2}+\eta(b,\sigma)-a\,Y_t\right)dt+\sqrt{Y_t}\,dB_t^{\mbq^{b,a,\sigma}}$$ with  a $\mathbb{Q}^{b,a,\sigma}$-Brownian motion \begin{equation}
	B_t^{\mbq^{b,a,\sigma}}:=-\eta(b,\sigma)\int_0^t\frac{1}{\sqrt{Y_s}}ds+B_t\,,\;0\le t\le T.
\end{equation} 
Then, \eqref{V:CIR} can be rewritten in the form
\begin{equation} \label{V:CIR:mart_ext}
	V(T\,;\,b,a,\sigma)=e^{-\lambda(b,a,\sigma) T}\sigma^{2(p\beta-\eta(b,\sigma))}\,\mathbb{E}^{\mathbb{Q}^{b,a,\sigma}}\bigl[Y_T^{p\beta-\eta(b,\sigma)}\bigr],
\end{equation}

The invariant density of the process $Y$ under $\mbq^{b,a,\sigma}$ is given as \cite[Theorem 5.1.10]{Pinsky1995}
\begin{equation}
	\psi(y)=\frac{(2a)^{\frac{2b}{\sigma^2}+2\eta(b,\sigma)}}{\Gamma\big(\frac{2b}{\sigma^2}+2\eta(b,\sigma)\big)}y^{\frac{2b}{\sigma^2}+2\eta(b,\sigma)-1}e^{-2ay}.
\end{equation}
Therefore, for any positive function $f$ on $(0,\infty)$ satisfying
\begin{equation} \label{inv:CIR}
	\int_{0}^{\infty}f(y)\psi(y)dy<\infty
\end{equation}
the expectation $\mathbb{E}^{\mathbb{Q}^{b,a,\sigma}}[f(Y_T)]$ converges to $\int_{0}^{\infty}f(y)\psi(y)dy$ as $T\to\infty$ \cite[Remark 4.2]{robertson2015large}. Clearly, $f(y):=y^{p\beta-\eta(b,\sigma)}$ can be shown to easily satisfy \eqref{inv:CIR}.
\vspace{3mm}

\begin{mycase}
	\case $\beta\geq 1$:
	By the comparison principle \cite[Chapter 5, Proposition 2.18]{karatzas1998brownian}, for every $(b,a,\sigma)\in[\underline{b},\overline{b}]\times[\underline{a},\overline{a}]\times[\underline{\sigma},\overline{\sigma}]$
	\begin{align}
		V(T\,;\,b,a,\sigma)&=\sigma^{2p\beta}\,\mathbb{E}^{\mbp^{b,a,\sigma}}\bigl[ Y_T^{p\beta}e^{-\frac{1}{2}p\beta(\beta-1)\int_0^T\frac{1}{Y_s}\,ds}\bigr] \\
		&\geq \underline{\sigma}^{2p\beta}\,\mathbb{E}^{\mbp^{\underline{b},\overline{a},\overline{\sigma}}}\bigl[ Y_T^{p\beta}e^{-\frac{1}{2}p\beta(\beta-1)\int_0^T\frac{1}{Y_s}\,ds}\bigr].
	\end{align}
	Hence,
	\begin{equation}
		v_T\geq e^{-p\,r(\beta-1)T}(\underline{\sigma}/\overline{\sigma})^{2p\beta}\,V(T\,;\,\underline{b},\overline{a},\overline{\sigma}).
	\end{equation}
	Applying the martingale extraction method to $V(T\,;\,\underline{b},\overline{a},\overline{\sigma})$ yields
	\begin{equation}
		V(T\,;\,\underline{b},\overline{a},\overline{\sigma})=e^{-\lambda(\underline{b},\overline{a},\overline{\sigma}) T}\overline{\sigma}^{2(p\beta-\eta(\underline{b},\overline{\sigma}))}\,\mathbb{E}^{\mathbb{Q}^{\underline{b},\overline{a},\overline{\sigma}}}\bigl[Y_T^{p\beta-\eta(\underline{b},\overline{\sigma})}\bigr],
	\end{equation}
	where $Y$ satisfies
	\begin{equation} \label{SDE:Q_dynamics:CIR:case1}			dY_t=\left(\frac{\underline{b}}{\overline{\sigma}^2}+\eta(\underline{b},\overline{\sigma})-\overline{a}\,Y_t\right)dt+\sqrt{Y_t}\,dB_t^{\mbq^{\underline{b},\overline{a},\overline{\sigma}}},\,Y_0=1/\overline{\sigma}^2,\quad \mbq^{\underline{b},\overline{a},\overline{\sigma}}\mbox{-a.s.}
	\end{equation}
	If $2\underline{b}/\overline{\sigma}^2+\eta(\underline{b},\overline{\sigma})+p\beta>0$, the expectation on the right-hand side of \eqref{SDE:Q_dynamics:CIR:case1} converges to some positive constant. Therefore,
	\begin{align}
		\liminf_{T\to\infty}\frac{1}{T}\log{v_T}&\geq -p\,r(\beta-1)-\overline{a}\,\eta(\underline{b},\overline{\sigma})+ \lim_{T\to\infty}\frac{1}{T}\left(\log{\underline{\sigma}^{2p\beta}\overline{\sigma}^{-2\eta(\underline{b},\overline{\sigma})}}+\log{\mathbb{E}^{\mathbb{Q}^{\underline{b},\overline{a},\overline{\sigma}}}\bigl[Y_T^{p\beta-\eta(\underline{b},\overline{\sigma})}\bigr]}\right)\\ &=-p\,r(\beta-1)-\overline{a}\,\eta(\underline{b},\overline{\sigma}). \label{conv_liminf:CIR}
	\end{align}
	
	Conversely, by definition of $v_T$, the inequality
	\begin{equation}
		e^{-p\,r(\beta-1)T}V(T\,;\,\underline{b},\overline{a},\overline{\sigma})\geq v_T
	\end{equation}
	holds for $\underline{b},\,\overline{a}$, and $\overline{\sigma}$, implying
	\begin{equation}
		-(p\,r(\beta-1)+\overline{a}\,\eta(\underline{b},\overline{\sigma}))T+2(p\beta-\eta(\underline{b},\overline{\sigma}))\log{\overline{\sigma}}+\log{\mathbb{E}^{\mathbb{Q}^{\underline{b},\overline{a},\overline{\sigma}}}\bigl[Y_T^{p\beta-\eta(\underline{b},\overline{\sigma})}\bigr]}\geq \log{v_T}.
	\end{equation}
	This leads to the inequality
	\begin{equation} \label{conv_limsup:CIR}
		-p\,r(\beta-1)-\overline{a}\,\eta(\underline{b},\overline{\sigma})\geq\limsup_{T\to\infty}\frac{1}{T}\log{v_T}.
	\end{equation}
	Thus, the two inequalities \eqref{conv_liminf:CIR} and \eqref{conv_limsup:CIR} yield
	\begin{equation}
		\lim_{T\to\infty}\frac{1}{T}\log{v_T}=-p\,r(\beta-1)-\overline{a}\,\eta(\underline{b},\overline{\sigma}),
	\end{equation}
	provided $2\underline{b}/\overline{\sigma}^2+\eta(\underline{b},\overline{\sigma})+p\beta>0$. \vspace{3mm}
	
	\case $0\leq \beta< 1$:
	We apply the comparison principle to \eqref{V:CIR:mart_ext}, such that
	\begin{equation}
		V(T\,;\,b,a,\sigma)\geq e^{-\underline{a}\,\eta(\overline{b},\underline{\sigma}) T}\sigma^{2(p\beta-\eta(\overline{b},\underline{\sigma}))}\,\mathbb{E}^{\mathbb{Q}^{\underline{b},\overline{a},\overline{\sigma}}}\bigl[Y_T^{p\beta-\eta(\overline{b},\underline{\sigma})}\bigr],
	\end{equation}
	Notably, $\eta<0$ for $0< \beta< 1$, establishing a difference from Case 1. Thus, we have
	\begin{equation}
		e^{-(p\,r(\beta-1)+\underline{a}\,\eta(\overline{b},\underline{\sigma})) T}\sigma^{2(p\beta-\eta(\overline{b},\underline{\sigma}))}\,\mathbb{E}^{\mathbb{Q}^{\underline{b},\overline{a},\overline{\sigma}}}\bigl[Y_T^{p\beta-\eta(\overline{b},\underline{\sigma})}\bigr]\leq v_T\leq e^{-p\,r(\beta-1)T}V(T\,;\,\overline{b},\underline{a},\underline{\sigma}).
	\end{equation}
	These inequalities produce the result
	\begin{equation}
		\lim_{T\to\infty}\frac{1}{T}\log{v_T}=-p\,r(\beta-1)-\underline{a}\,\eta(\overline{b},\underline{\sigma}).
	\end{equation}

	\case $\beta<0$:
	In this case, $p\beta-\eta(b,\sigma)<0$ for all $(b,\sigma)\in[\underline{b},\overline{b}]\times[\underline{\sigma},\overline{\sigma}]$. Hence, the comparison principle leads to
	\begin{equation}
		e^{-(p\,r(\beta-1)+\overline{a}\,\eta(\underline{b},\overline{\sigma})) T}\sigma^{2(p\beta-\eta(\underline{b},\overline{\sigma}))}\,\mathbb{E}^{\mathbb{Q}^{\overline{b},\underline{a},\underline{\sigma}}}\bigl[Y_T^{p\beta-\eta(\underline{b},\overline{\sigma})}\bigr]\leq v_T\leq e^{-p\,r(\beta-1)T}V(T\,;\,\underline{b},\overline{a},\overline{\sigma}),
	\end{equation}
	Finally,
	\begin{equation}
		\lim_{T\to\infty}\frac{1}{T}\log{v_T}=-p\,r(\beta-1)-\overline{a}\,\eta(\underline{b},\overline{\sigma}).
	\end{equation}
	
\end{mycase}

The obtained results are summarized as follows.
\begin{proposition} \label{prop:CIR}
	Let $0<\underline{a}\leq\overline{a},\,0<\underline{\sigma}\leq\overline{\sigma},\,\overline{\sigma}^2<\underline{b}\leq\overline{b}$ and $X^{\alpha}$ be the CIR process \eqref{SDE:CIR} with set of parameters $\alpha=(b,a,\sigma)$ ranging over $[\underline{\alpha},\overline{\alpha}]=[\underline{b},\overline{b}]\times[\underline{a},\overline{a}]\times[\underline{\sigma},\overline{\sigma}]$.
	Then, the long-term growth rate of the worst-case expected utility of the LETF $L^{\alpha}=(L_t^{\alpha})_{t\geq 0}$, with the reference process $X$, is given by
	\begin{equation}
		\lim_{T\to\infty}\frac{1}{T}\log{\inf_{\alpha\in[\underline{\alpha},\overline{\alpha}]}\mathbb{E}^{\mbp^{\alpha}}[L_T^p]}=-p\,r(\beta-1)-a^*(\beta)\eta(b^*(\beta),\sigma^*(\beta),\beta),
	\end{equation}
	provided $2\underline{b}/\overline{\sigma}^2+\eta+p\beta>0$, where
	\begin{equation}
		\eta(b,\sigma,\beta)=-\Big(\frac{b}{\sigma^2}-\frac{1}{2}\Big)+\sqrt{\Big(\frac{b}{\sigma^2}-\frac{1}{2}\Big)^2+p\beta(\beta-1)},
	\end{equation}
	\begin{equation}
		b^*(\beta)=\begin{cases}\underline{b}, &  \beta\geq 1,\,\beta<0 \\ \overline{b}, & 0\leq\beta<1\end{cases},\,a^*(\beta)=\begin{cases}\overline{a}, &  \beta\geq 1,\,\beta<0 \\ \underline{a}, & 0\leq\beta<1\end{cases},\,\sigma^*(\beta)=\begin{cases}\overline{\sigma}, &  \beta\geq 1,\,\beta<0 \\ \underline{\sigma}, & 0\leq\beta<1\end{cases}.
	\end{equation}
\end{proposition}
Proposition \ref{prop:CIR} implies the $\beta$ dependency of the parameters achieving robust long-term growth rate. Thus, to obtain an optimal $\beta^*\in[\underline{\beta},\overline{\beta}]$ maximizing the robust long-term growth rate, finding $\beta^*$ for each case and comparing them are necessary. Therefore, we define a function $\Lambda$ of $\beta$ as
$$\Lambda(\beta):=-p\,r(\beta-1)-a^*(\beta)\eta(b^*(\beta),\sigma^*(\beta),\beta).$$

\begin{mycase}
	\case $\beta\geq 1$:
	The first derivative of $\Lambda$ is given by
	\begin{equation}
		\Lambda'(\beta)=-p\,r-\frac{\overline{a}p(2\beta-1)}{2\sqrt{(\frac{\underline{b}}{\overline{\sigma}^2}-\frac{1}{2})^2+p\beta(\beta-1)}},
	\end{equation}
	which is definitely negative. Thus, $\Lambda$ achieves its maximum at $\beta=1$ on $[1,5]$. However, $\beta=1$ cannot be an optimal leverage ratio because
	\[\Lambda(1)=0<p\,r=\Lambda(0). \]
	
	\case $0\leq \beta<1$:
	If $r^p\geq\overline{a}^2$, or both $r^2p<\overline{a}^2$ and $\frac{2\underline{b}}{\overline{\sigma}^2}-1\geq\frac{\overline{a}}{r}$, then $\Lambda'(\beta)<0$ on $[0,1)$; hence, $\beta^*=0$. Otherwise, $\beta^*=\frac{1}{2}\left(1-\sqrt{\frac{(\frac{2\underline{b}}{\overline{\sigma}^2}-1)^2-p}{(\frac{\overline{a}}{r})^2-p}}\right).$ \vspace{3mm}
	
	\case $\beta<0$:
	If $r^p\geq\underline{a}^2$, or both $r^2p<\underline{a}^2$ and $\frac{2\overline{b}}{\underline{\sigma}^2}-1\leq\frac{\underline{a}}{r}$, then $\Lambda'(\beta)<0$ on $[-5,0)$; hence, $\beta^*=-5$. Otherwise, $\beta^*=\frac{1}{2}\left(1-\sqrt{\frac{(\frac{2\overline{b}}{\underline{\sigma}^2}-1)^2-p}{(\frac{\underline{a}}{r})^2-p}}\right).$
\end{mycase} \vspace{3mm}

Unfortunately, the overall optimal value of $\beta^*$ depends on the relationships among the parameters; hence, $\beta^*$ may vary according to the uncertainty set. We summarize the results in Table \ref{table:CIR}.

\begin{table}[h]
	\centering
	\begin{tabular}{|c|c|c|c|}\hline
		\multicolumn{3}{|c|}{Parameter relationship} & Candidates for $\beta^*$ \\ \hline
		\multicolumn{3}{|c|}{$p\geq\frac{\overline{a}^2}{r^2}$} & $0,\,\underline{\beta}$ \\ \cline{1-3}
		$\frac{\underline{a}^2}{r^2}\leq p<\frac{\overline{a}^2}{r^2}$& \multicolumn{2}{|c|}{$\frac{2\underline{b}}{\overline{\sigma}^2}-1\geq\frac{\overline{a}}{r}$} &  \\[1mm] \cline{2-4}
		& \multicolumn{2}{|c|}{$\frac{2\underline{b}}{\overline{\sigma}^2}-1<\frac{\overline{a}}{r}$} & $\frac{1}{2}\Big(1-\sqrt{\frac{(\frac{2\underline{b}}{\overline{\sigma}^2}-1)^2-p}{(\frac{\overline{a}}{r})^2-p}}\Big),\,\underline{\beta} $ \\ \hline
		$p<\frac{\underline{a}^2}{r^2}$ & \multicolumn{2}{|c|}{$\frac{2\underline{b}}{\overline{\sigma}^2}-1\geq\frac{\overline{a}}{r}$} & $0,\,\frac{1}{2}\Big(1-\sqrt{\frac{(\frac{2\overline{b}}{\underline{\sigma}^2}-1)^2-p}{(\frac{\underline{a}}{r})^2-p}}\Big)$ \\ \cline{2-4}
		& $\frac{2\underline{b}}{\overline{\sigma}^2}-1<\frac{\overline{a}}{r}$ & $\frac{2\overline{b}}{\underline{\sigma}^2}-1>\frac{\underline{a}}{r}$ & $\frac{1}{2}\Big(1-\sqrt{\frac{(\frac{2\underline{b}}{\overline{\sigma}^2}-1)^2-p}{(\frac{\overline{a}}{r})^2-p}}\Big),\,\frac{1}{2}\Big(1-\sqrt{\frac{(\frac{2\overline{b}}{\underline{\sigma}^2}-1)^2-p}{(\frac{\underline{a}}{r})^2-p}}\Big)$ \\ \cline{3-4}
		& & $\frac{2\overline{b}}{\underline{\sigma}^2}-1\leq\frac{\underline{a}}{r}$ & $\frac{1}{2}\Big(1-\sqrt{\frac{(\frac{2\underline{b}}{\overline{\sigma}^2}-1)^2-p}{(\frac{\overline{a}}{r})^2-p}}\Big),\,\underline{\beta}$ \\ \hline
	\end{tabular}\caption{Table of results under the CIR model.}\label{table:CIR}\end{table}

\subsection{3/2 model}
\label{sec:3/2_model}
A 3/2 model is a positive non-affine model of the form
\begin{equation}\label{SDE:3/2}
	dX_t=(b-aX_t)X_t\,dt+\sigma X_t^{3/2}\,dB_t\,,
\end{equation}
with $b,\,\sigma>0$, and $a>-\sigma^2/2$. This model is a suggested alternative to the CIR model for stochastic volatility (\citet{carr2007new} and \citet{drimus2012options}) and short interest rates (\citet{ahn1999parametric}). In this model, we set $\underline{\alpha}=(\underline{b},\underline{a},\underline{\sigma}),\,\overline{\alpha}=(\overline{b},\overline{a},\overline{\sigma}),\,[\underline{\alpha},\overline{\alpha}]=[\underline{b},\overline{b}]\times[\underline{a},\overline{a}]\times[\underline{\sigma},\overline{\sigma}]$ with $\underline{b},\underline{a},\underline{\sigma}>0$. The worst-case expected utility $v_T$ is given by
\begin{align}
	v_T&=e^{-p\,r(\beta-1)T}\inf_{\alpha\in[\underline{\alpha},\overline{\alpha}]}V(T\,;\,\alpha) \\
	&=e^{-p\,r(\beta-1)T}\inf_{\alpha\in[\underline{\alpha},\overline{\alpha}]}\mathbb{E}^{\mbp^{\alpha}}\bigl[X^{p\beta} e^{-\frac{1}{2}p\beta(\beta-1)\int_0^T\sigma^2X_s\,ds}\bigr].
\end{align}

A process $Y$ defined as $Y:=\sigma^2X$ can be considered as another 3/2 process satisfying
\begin{equation}
	dY_t=\left(b-\frac{a}{\sigma^2}Y_t\right)Y_tdt+Y_t^{3/2}\,dB_t,\quad Y_0=\sigma^2.
\end{equation}
Then, $V$ can be written in terms of $Y$ as
\begin{equation}
	V(T\,;\,\alpha)=\sigma^{-2p\beta}\,\mathbb{E}^{\mbp^{\alpha}}\bigl[ Y_T^{p\beta}e^{-\frac{1}{2}p\beta(\beta-1)\int_0^T Y_s\,ds}\bigr].
\end{equation}
The eigenpair problem for the infinitesimal generator of $Y$ is expressed as
\begin{equation}
	-\lambda\phi(y)=\frac{1}{2}y^3\phi''(y)+\left(b-\frac{a}{\sigma^2}y\right)y\phi'(y)-\frac{1}{2}p\beta(\beta-1)y\phi(y)
\end{equation}
with one of its solutions as
\begin{equation} \label{eigpair:3/2}
	(\lambda(b,a,\sigma),\phi(y))=(b\,\eta(a,\sigma),\,y^{-\eta(a,\sigma)}),
\end{equation}
where
\begin{equation} \label{eta:3/2}
	\eta(a,\sigma)=-\Big(\frac{a}{\sigma^2}+\frac{1}{2}\Big)+\sqrt{\Big(\frac{a}{\sigma^2}+\frac{1}{2}\Big)^2+p\beta(\beta-1)}.
\end{equation}
Here, $\eta$ is real because $a/\sigma^2>0$. The martingale extraction method applied to $V(T\,;\,b,a,\sigma)$ yields
\begin{equation} \label{V:3/2:mart_ext}
	V(T\,;\,b,a,\sigma)=e^{-\lambda(b,a,\sigma) T}\sigma^{-2(p\beta+\eta(a,\sigma))}\,\mathbb{E}^{\mathbb{Q}^{b,a,\sigma}}\bigl[Y_T^{p\beta+\eta(a,\sigma)}\bigr],
\end{equation}
where $\mathbb{Q}^{b,a,\sigma}$ is defined on $\mathcal{F}_T$ by
\begin{equation} \label{RN:3/2}
	\frac{d\mathbb{Q}^{b,a,\sigma}}{d\mathbb{P}}\biggl|_{\mathcal{F}_T}=\left(\frac{Y_T}{y}\right)^{-\eta(a,\sigma)}\exp\left\{\lambda(b,a,\sigma) T-\frac{1}{2}p\beta(\beta-1)\int_0^TY_s\,ds\right\},
\end{equation}
under which $Y$ follows
\begin{equation} \label{SDE:Q_dynamics:3/2}
	dY_t=\left(b-\left(\frac{a}{\sigma^2}+\eta(a,\sigma)\right)Y_t\right)Y_t\,dt+Y_t^{3/2}\,dB_t^{\mbq^{b,a,\sigma}},
\end{equation}
with $\mbq^{b,a,\sigma}$-Brownian motion
\begin{equation}
	B_t^{\mbq^{b,a,\sigma}}:=\eta(a,\sigma)\int_0^t\sqrt{Y_s}ds+B_t.
\end{equation}
The Radon--Nikodym derivative \eqref{RN:3/2} is well-defined using an argument similar to that used in the CIR model.

The invariant density of the process $Y$ under $\mbq^{b,a,\sigma}$ is given by
\begin{equation}
	\psi(y)=\frac{(2b)^{\frac{2a}{\sigma^2}+2\eta(b,\sigma)}}{\Gamma\big(\frac{2a}{\sigma^2}+2\eta(b,\sigma)+2\big)}y^{-\frac{2a}{\sigma^2}-2\eta(b,\sigma)-3}e^{-\frac{2b}{y}}.
\end{equation}
Thus, we can show that
\begin{equation} \label{inv:3/2}
	\int_0^{\infty}y^{p\beta+\eta(a,\sigma)}\psi(y)dy<\infty,
\end{equation}
and, consequently, the expectation on the right-hand side of \eqref{V:3/2:mart_ext} converges to \eqref{inv:3/2}.
Additionally, the comparison principle is applicable to $Y$ as well, given that $\frac{1}{Y}$ is a CIR process. \vspace{3mm}

\begin{mycase}
	\case $\beta\geq 1$:
	The comparison principle is applied to \eqref{V:3/2:mart_ext}, such that
	\begin{equation}
		V(T\,;\,b,a,\sigma)\geq e^{-\overline{b}\,\eta(\underline{a},\overline{\sigma}) T}\overline{\sigma}^{-2(p\beta+\eta(\underline{a},\overline{\sigma}))}\,\mathbb{E}^{\mathbb{Q}^{\underline{b},\overline{a},\underline{\sigma}}}\bigl[Y_T^{p\beta+\eta(\underline{a},\overline{\sigma})}\bigr],
	\end{equation}
	Thus, we have
	\begin{equation}
		e^{-(p\,r(\beta-1)+\overline{b}\,\eta(\underline{a},\overline{\sigma})) T}\overline{\sigma}^{-2(p\beta+\eta(\underline{a},\overline{\sigma}))}\,\mathbb{E}^{\mathbb{Q}^{\underline{b},\overline{a},\underline{\sigma}}}\bigl[Y_T^{p\beta+\eta(\underline{a},\overline{\sigma})}\bigr]\leq v_T\leq e^{-p\,r(\beta-1)T}V(T\,;\,\overline{b},\underline{a},\overline{\sigma}).
	\end{equation}
	Therefore,
	\begin{equation}
		\lim_{T\to\infty}\frac{1}{T}\log{v_T}=-p\,r(\beta-1)-\overline{b}\,\eta(\underline{a},\overline{\sigma}).
	\end{equation}
	
	\case $0\leq\beta<1$:
	Since $\eta<0$ for $0< \beta< 1$, $p\beta+\eta(a,\sigma)<0$ for all $(a,\sigma)\in[\underline{a},\overline{a}]\times[\underline{\sigma},\overline{\sigma}]$. Consequently,
	\begin{equation}
		\lim_{T\to\infty}\frac{1}{T}\log{v_T}=-p\,r(\beta-1)-\underline{b}\,\eta(\overline{a},\underline{\sigma})
	\end{equation}
	is obtained, which is the complete opposite of what we observed in Case 1. \vspace{3mm}
	
	\case $\beta<0$:
	For every $\alpha=(b,a,\sigma)\in[\underline{a},\overline{a}]$,
	\begin{align}
		V(T\,;\,b,a,\sigma)&=\sigma^{-2p\beta}\,\mathbb{E}^{\mbp^{b,a,\sigma}}\bigl[ Y_T^{p\beta}e^{-\frac{1}{2}p\beta(\beta-1)\int_0^T Y_s\,ds}\bigr] \\
		&\geq \underline{\sigma}^{-2p\beta}\,\mathbb{E}^{\mbp^{\overline{b},\underline{a},\overline{\sigma}}}\bigl[ Y_T^{p\beta}e^{-\frac{1}{2}p\beta(\beta-1)\int_0^T Y_s\,ds}\bigr] \\
		&=(\underline{\sigma}/\overline{\sigma})^{-2p\beta}\,V(T\,;\,\overline{b},\underline{a},\overline{\sigma}).
	\end{align}
	Therefore
	\begin{equation}
		v_T\geq e^{-p\,r(\beta-1)T}(\underline{\sigma}/\overline{\sigma})^{-2p\beta}\,V(T\,;\,\overline{b},\underline{a},\overline{\sigma}).
	\end{equation}
	The martingale extraction method applied to $V(T\,;\,\overline{b},\underline{a},\overline{\sigma})$ yields
	\begin{equation}
		V(T\,;\,\overline{b},\underline{a},\overline{\sigma})=e^{-\lambda(\overline{b},\underline{a},\overline{\sigma}) T}\overline{\sigma}^{-2(p\beta+\eta(\underline{a},\overline{\sigma}))}\,\mathbb{E}^{\mathbb{Q}^{\overline{b},\underline{a},\overline{\sigma}}}\bigl[Y_T^{p\beta+\eta(\underline{a},\overline{\sigma})}\bigr],
	\end{equation}
	where $Y$ follows
	\begin{equation} \label{SDE:Q_dynamics:3/2:case3}
		dY_t=\left(\overline{b}-\left(\frac{\underline{a}}{\overline{\sigma}^2}+\eta(\underline{a},\overline{\sigma})\right)Y_t\right)Y_t\,dt+Y_t^{3/2}\,dB_t^{\mbq},\,Y_0=\overline{\sigma}^2,\quad\mbq^{\overline{b},\underline{a},\overline{\sigma}}\mbox{-a.s.}
	\end{equation}
	If $2(\underline{a}/\overline{\sigma}^2+1)+\eta(\underline{a},\overline{\sigma})-p\beta>0,\,\,\dfrac{1}{T}\log{V(T\,;\,\overline{b},\underline{a},\overline{\sigma})}$ converges to $-\lambda(\overline{b},\underline{a},\overline{\sigma})=-\overline{b}\,\eta(\underline{a},\overline{\sigma})$ as $T\to\infty$. Therefore,
	\begin{align}
		\liminf_{T\to\infty}\frac{1}{T}\log{v_T}&\geq -p\,r(\beta-1)-\overline{b}\,\eta(\underline{a},\overline{\sigma})+ \liminf_{T\to\infty}\frac{1}{T}\left(\log{\frac{\underline{\sigma}^{-2p\beta}}{\overline{\sigma}^{-2\eta(\underline{a},\overline{\sigma})}}}+\log{\mathbb{E}^{\mathbb{Q}^{\overline{b},\underline{a},\overline{\sigma}}}\bigl[Y_T^{p\beta+\eta(\underline{a},\overline{\sigma})}\bigr]}\right)\\ &=-p\,r(\beta-1)-\overline{b}\,\eta(\underline{a},\overline{\sigma}).
	\end{align}
	
	Conversely, by definition of $v_T$, the inequality
	\begin{equation}
		e^{-p\,r(\beta-1)T}V(T\,;\,\overline{b},\underline{a},\overline{\sigma})\geq v_T
	\end{equation}
	holds for $\overline{b},\,\underline{a}$, and $\overline{\sigma}$, implying
	\begin{equation}
		-(p\,r(\beta-1)+\overline{b}\,\eta(\underline{a},\overline{\sigma}))T-2(p\beta+\eta(\underline{a},\overline{\sigma}))\log{\overline{\sigma}}+\log{\mathbb{E}^{\mathbb{Q}^{\overline{b},\underline{a},\overline{\sigma}}}\bigl[Y_T^{p\beta+\eta(\underline{a},\overline{\sigma})}\bigr]}\geq \log{v_T}.
	\end{equation}
	This leads to the inequality
	\begin{equation}
		-p\,r(\beta-1)-\overline{b}\,\eta(\underline{a},\overline{\sigma})\geq\limsup_{T\to\infty}\frac{1}{T}\log{v_T}.
	\end{equation}
	Combining the two aforementioned inequalities yields
	\begin{equation}
		\lim_{T\to\infty}\frac{1}{T}\log{v_T}=-p\,r(\beta-1)-\overline{b}\,\eta(\underline{a},\overline{\sigma}).
	\end{equation}
	provided that $2(\underline{a}/\overline{\sigma}^2+1)+\eta(\underline{a},\overline{\sigma})-p\beta>0$.
	
\end{mycase} \vspace{3mm}

The obtained results have been summarized as follows.
\begin{proposition} \label{prop:3/2}
	Let $0<\underline{b}\leq\overline{b},\,0<\underline{\sigma}\leq\overline{\sigma},\,0<\underline{a}\leq\overline{a}$ and $X^{\alpha}$ be the 3/2 process \eqref{SDE:3/2} with set of parameters $\alpha=(b,a,\sigma)$ ranging over $[\underline{\alpha},\overline{\alpha}]=[\underline{b},\overline{b}]\times[\underline{a},\overline{a}]\times[\underline{\sigma},\overline{\sigma}]$.
	Then the long-term growth rate of the worst-case expected utility of the LETF $L^{\alpha}=(L_t^{\alpha})_{t\geq 0}$, with the reference process $X^{\alpha}$ is given by
	\begin{equation}
		\lim_{T\to\infty}\frac{1}{T}\log{\inf_{\alpha\in[\underline{\alpha},\overline{\alpha}]}\mathbb{E}^{\mbp^{\alpha}}[L_T^p]}=-p\,r(\beta-1)-b^*(\beta)\,\eta(a^*(\beta),\sigma^*(\beta),\beta),
	\end{equation}
	provided $2(\underline{a}/\overline{\sigma}^2+1)+\eta(\underline{a},\overline{\sigma})-p\beta>0$, where
	\begin{equation}
		\eta(a,\sigma,\beta)=-\Big(\frac{a}{\sigma^2}+\frac{1}{2}\Big)+\sqrt{\Big(\frac{a}{\sigma^2}+\frac{1}{2}\Big)^2+p\beta(\beta-1)},
	\end{equation}
	\begin{equation}
		b^*(\beta)=\begin{cases}\overline{b}, &  \beta\geq 1,\,\beta<0 \\ \underline{b}, & 0\leq\beta<1\end{cases},\,a^*(\beta)=\begin{cases}\underline{a}, &  \beta\geq 	1,\,\beta<0 \\ \overline{a}, & 0\leq\beta<1\end{cases},\,\sigma^*(\beta)=\begin{cases}\overline{\sigma}, &  \beta\geq 1,\,\beta<0 \\ \underline{\sigma}, & 0\leq\beta<1\end{cases}.
	\end{equation}
\end{proposition}

Proposition \ref{prop:3/2} is similar to Proposition \ref{prop:CIR}; thus the optimal leverage ratio candidates according to the parameter relationship can be summarized, as listed in Table \ref{table:3/2}.

\begin{table}[h]
	\begin{center}\begin{tabular}{|c|c|c|c|} \hline
			\multicolumn{3}{|c|}{Parameter relationship} & Candidates for $\beta^*$ \\ \hline
			\multicolumn{3}{|c|}{$p\geq\frac{\overline{b}^2}{r^2}$} & $0,\,\underline{\beta}$ \\ \hline
			$\frac{\underline{b}^2}{r^2}\leq p<\frac{\overline{b}^2}{r^2}$& \multicolumn{2}{|c|}{$\frac{2\underline{a}}{\overline{\sigma}^2}+1>\frac{\overline{b}}{r}$} & $0,\,\frac{1}{2}\Big(1-\sqrt{\frac{(\frac{2\underline{a}}{\overline{\sigma}^2}+1)^2-p}{(\frac{\overline{b}}{r})^2-p}}\Big)$ \\[1mm] \cline{2-4}
			& \multicolumn{2}{|c|}{$\frac{2\underline{a}}{\overline{\sigma}^2}+1\leq\frac{\overline{b}}{r}$} & $0,\,\underline{\beta} $ \\ \hline
			$p<\frac{\underline{b}^2}{r^2}$ & \multicolumn{2}{|c|}{$\frac{2\underline{a}}{\overline{\sigma}^2}+1>\frac{\overline{b}}{r}$} & $0,\,\frac{1}{2}\Big(1-\sqrt{\frac{(\frac{2\underline{a}}{\overline{\sigma}^2}+1)^2-p}{(\frac{\overline{b}}{r})^2-p}}\Big)$ \\ \cline{2-4}
			& $\frac{2\underline{a}}{\overline{\sigma}^2}+1\leq\frac{\overline{b}}{r}$ & $\frac{2\overline{a}}{\underline{\sigma}^2}+1\geq\frac{\underline{b}}{r}$ & $0,\,\underline{\beta}$ \\ \cline{3-4}
			& & $\frac{2\overline{a}}{\underline{\sigma}^2}+1<\frac{\underline{b}}{r}$ & $\frac{1}{2}\Big(1-\sqrt{\frac{(\frac{2\overline{a}}{\underline{\sigma}^2}+1)^2-p}{(\frac{\underline{b}}{r})^2-p}}\Big),\,\underline{\beta}$ \\ \hline
		\end{tabular}	\caption{Table of results under the 3/2 model.}\label{table:3/2}
	\end{center}\end{table}

\section{Uncertainties on stochastic volatility reference}
\label{sec:SV}

In this section, we maintain the assumption that the interest rate is constant: $r_t\equiv r$ and the initial value of the reference process is 1. The reference process $X$ considered in this section expressed as
\begin{align}
	&dX_t=\mu X_t\,dt+\sqrt{\nu_t}\,X_t\,dW_t\;,\\
	&d\nu_t=b(\nu_t)\,dt+\sigma(\nu_t)\,dB_t,\quad\nu_0>0.
\end{align}
Assuming that uncertainties lie in $\mu$, instantaneous correlation coefficient $\rho$ between $W$ and $B$, and the set of parameters $\tilde{\alpha}$ of $\nu$, we have $\alpha=(\mu,\rho,\tilde{\alpha})$ and $[\underline{\alpha},\overline{\alpha}]=[\underline{\mu},\overline{\mu}]\times[\underline{\rho},\overline{\rho}]\times[\underline{\tilde{\alpha}},\overline{\tilde{\alpha}}]$ where $-1\leq\underline{\rho}\leq\overline{\rho}\leq 1$. Then, the dynamics of $X$ and $\nu$ can be rewritten in the form that emphasizes dependence on parameters $\alpha$, such that
\begin{align}
	&dX_t^{\alpha}=\mu X_t^{\alpha}\,dt+\sqrt{\nu_t}X_t^{\alpha}\,dW_t^{\rho}\;,\\
	&d\nu_t^{\alpha}=b(\nu_t^{\alpha};\alpha)\,dt+\sigma(\nu_t^{\alpha};\alpha)\,dB_t^{\rho}\;,
\end{align}
with $\langle W^{\rho},B^{\rho}\rangle_t=\rho\,t$. Correspondingly,
\begin{align}
	(L_t^{\alpha})^p&=e^{p(r+\beta(\mu-r))t+p\beta\int_0^t\sqrt{\nu_s^{\alpha}}dW_s^{\rho}-\frac{1}{2}p\beta^2\int_0^t\nu_s^{\alpha}ds}\\
	&=e^{p(r+\beta(\mu-r))t-\frac{1}{2}p(1-p)\beta^2\int_0^t\nu_s^{\alpha}ds}\mathcal{E}\!\left(p\beta\!\int_0^\cdot\sqrt{\nu_s^{\alpha}}dW_s^{\rho}\right)_t,\quad t\geq 0.
\end{align}
Fixing $T>0$ and assuming that the exponential martingale term is a true martingale for $0\leq t\leq T$, for a probability measure $\hat{\mbp}^{\alpha}$ defined on $\mathcal{F}_T$ by
\begin{equation} \label{measure:SV}
	\frac{d\hat{\mathbb{P}}^{\alpha}}{d\mbp}\biggl|_{\mathcal{F}_T}=\mathcal{E}\!\left(p\beta\!\int_0^\cdot\sqrt{\nu_s^{\alpha}}dW_s^{\rho}\right)_T,
\end{equation}
the expected utility of an investor holding the LETF can be expressed as
\begin{equation} \label{temp:SV}
	\mathbb{E}^{\mbp^{\alpha}}[L_T^p]=\mathbb{E}^{\hat{\mathbb{P}}^{\alpha}}[e^{p(r+\beta(\mu-r))T-\frac{1}{2}p(1-p)\beta^2\int_0^T\nu_s^{\alpha}ds}],\quad T\geq 0.
\end{equation}
By Girsanov's theorem, the $\hat{\mathbb{P}}^{\alpha}$-dynamics of $\nu^{\alpha}$ is given by
\begin{equation}
	d\nu_t^{\alpha}=\left(b(\nu_t^{\alpha};\alpha)+p\beta\rho\sqrt{\nu_t^{\alpha}}\sigma(\nu_t^{\alpha};\alpha)\right)dt+\sigma(\nu_t^{\alpha};\alpha)\,d\hat{B}_t^{\alpha},\quad 0\leq t\leq T,
\end{equation}
with a $\hat{\mathbb{P}}^{\alpha}$-Brownian motion
\begin{equation}
	\hat{B}_t^{\alpha}:=-p\beta\rho\int_0^t\sqrt{\nu_s^{\alpha}}ds+B_t^{\rho},\quad 0\leq t\leq T.
\end{equation}
The inequality is straightfoward from \eqref{temp:SV} that
\begin{equation} \label{ineqn:SV}
	\mathbb{E}^{\mbp^{\alpha}}[L_T^p]\geq e^{p(r+\beta(\mu^{*}-r))T}\mathbb{E}^{\hat{\mbp}^{\alpha}}[e^{-\frac{1}{2}p(1-p)\beta^2\int_0^T\nu_s^{\alpha}ds}],\quad T\geq 0,
\end{equation}
for every $\alpha\in[\underline{\alpha},\overline{\alpha}]$, where
\begin{equation}
	\mu^{*}=\begin{cases}
		\underline{\mu}, &  \beta\geq 0 \\ \overline{\mu}, & \beta<0.
	\end{cases}
\end{equation}
Similar to the previous section, we set
\begin{align}
	V(T\,;\,\alpha)&:=\mathbb{E}^{\hat{\mbp}^{\alpha}}[e^{-\frac{1}{2}p(1-p)\beta^2\int_0^T\nu_s^{\alpha}ds}],\\
	v_T&:=\inf_{\alpha\in[\underline{\alpha},\overline{\alpha}]}\mathbb{E}^{\mbp^{\alpha}}[L_T^p].
\end{align}
As shown in \eqref{eqn:V,v:reference}, the following can be shown.
\begin{equation} \label{eqn:V,v:SV}
	v_T= e^{p(r+\beta(\mu^{*}-r))T}\inf_{\alpha\in[\underline{\alpha},\overline{\alpha}]}\!V(T\,;\,\alpha)
\end{equation}
Then, the comparison principle and martingale extraction method can be applied to $\nu^{\alpha}$ and $V(T\,;\,\alpha)$, respectively.

\subsection{Heston model}
\label{sec:Heston}
The Heston stochastic volatility model was suggested by \citet{heston1993closed} to overcome the shortcomings of the Black--Scholes model, which assumes volatility to be constant over time. Suppose that the reference follows the Heston model
\begin{equation}\label{eqn:Heston}
	\begin{aligned}
		&dX_t=\mu X_t\, dt+ \sqrt{\nu_t}\,X_t\,dW_t\;,\\
		&d\nu_t=(b-a\,\nu_t)\,dt+\sigma\sqrt{\nu_t}\,dB_t\;,
	\end{aligned}
\end{equation}
where $W_t$ and $B_t$ are two correlated  Brownian motions with  $\langle W,B\rangle_t=\rho\,t$  and correlation parameter  $\rho \in [\underline{\rho},\overline{\rho}]$. Then, $\alpha=(\mu,\rho,b,a,\sigma)$ and $[\underline{\alpha},\overline{\alpha}]=[\underline{\mu},\overline{\mu}]\times[\underline{\rho},\overline{\rho}]\times[\underline{b},\overline{b}]\times[\underline{a},\overline{a}]\times[\underline{\sigma},\overline{\sigma}]$, with $\underline{\mu},\underline{a},\underline{\sigma}>0$  and $\underline{b}>\overline{\sigma}^2/2$. Assuming $\underline{a}-p|\beta|\overline{\sigma}>0$ so that $a-p\beta\rho\,\sigma>0$ for every $(a,\rho,\sigma)\in[\underline{a},\overline{a}]\times[-1,1]\times[\underline{\sigma},\overline{\sigma}]$, the SDE
\begin{equation}
	d\nu_t=(b-(a-p\beta\rho\,\sigma)\nu_t)dt+\sigma\sqrt{\nu_t}\,dB_t,\quad \nu_0>0,
\end{equation}
has a unique strong solution, ensuring that a probability measure $\hat{\mbp}^{\alpha}$ can be defined on $\mathcal{F}_T$ for each $T\geq0$ by \eqref{measure:SV}. Hence, the inequality \eqref{ineqn:SV} holds, and under $\hat{\mbp}^{\alpha},\,\nu$ follows
\begin{equation}
	d\nu_t=(b-(a-p\beta\rho\,\sigma)\nu_t)dt+\sigma\sqrt{\nu_t}\,d\hat{B}_t^{\alpha}
\end{equation}
with a $\hat{\mbp}^{\alpha}$-Brownian motion
$$\hat{B}_t^{\alpha}=-p\beta\rho\int_0^t\sqrt{\nu_s}\,ds+B_t.$$
Moreover, by applying the comparison principle to $\nu$, we have
\begin{equation}
	\mathbb{E}^{\mbp^{\alpha}}[L_T^p]\geq e^{p(r+\beta(\mu^{*}-r))T}\mathbb{E}^{\hat{\mbp}^{\mu^{*},\rho^*,\overline{b},\underline{a},\sigma}}[e^{-\frac{1}{2}p(1-p)\beta^2\int_0^T\nu_sds}],\quad T\geq 0,
\end{equation}
for every $\alpha=(\mu,\rho,b,a,\sigma)$, where
\begin{equation}
	\rho^{*}=\begin{cases}
		\overline{\rho}, &  \beta\geq 0 \\ \underline{\rho}, & \beta<0.
	\end{cases}
\end{equation}
So far, $\sigma$ has not been specified. For notational simplicity, set
\begin{equation}
	\alpha^{*}(\sigma):=(\mu^{*},\rho^*,\overline{b},\underline{a},\sigma).
\end{equation}

Under the measure $\hat{\mbp}^{\alpha^{*}(\sigma)}$, the eigenpair problem for the infinitesimal generator of $\nu$ is expressed as
$$-\lambda\phi(\nu)=\frac{1}{2}\sigma^2\nu\,\phi''(\nu)+(\overline{b}-(\underline{a}-p\beta\rho^*\sigma) \nu)\,\phi'(\nu)-\frac{1}{2}p(1-p)\beta^2\nu\,\phi(\nu)\;.$$
One solution pair is given by
\begin{equation*}
	(\lambda(\sigma),\phi_{\sigma}(v))=\left(\overline{b}\,\eta(\sigma),e^{-\eta(\sigma) v}\right) \,,
\end{equation*}
where
\begin{equation}
	\eta(\sigma)=\frac{1}{\sigma^2}\left(\sqrt{\left(\underline{a}-p\beta\rho^*\sigma\right)^2+p(1-p)\beta^2\sigma^2}-\left(\underline{a}-p\beta\rho^*\sigma\right)\right).
\end{equation}
Arguments similar to those used in Section \ref{sec:Uncertain_ref} show that $\hat{\mbq}^{\alpha^*\!(\sigma)}$ defined by
$$\frac{d\hat{\mathbb{Q}}^{\alpha^{*}\!(\sigma)}}{d\hat{\mathbb{P}}^{\alpha^{*}\!(\sigma)}}\biggl|_{\mathcal{F}_T}=\exp{\left\{\lambda(\sigma) T-\frac{1}{2}p\beta(\beta-1)\int_0^T\nu_sds-\eta(\sigma)\nu_T+\eta(\sigma)\nu_0\right\}}$$
is a probability measure on $\mathcal{F}_T$ under which the process $\nu$ satisfies
\begin{equation}
	d\nu_t=(\overline{b}-\sqrt{(\underline{a}-p\beta\rho^*\sigma)^2+p(1-p)\beta^2\sigma^2}\nu_t)dt+\sigma\sqrt{\nu_t}dB_t^{\hat{\mbq}^{\alpha^*\!(\sigma)}},\quad 0\leq t\leq T,
\end{equation}
where $B^{\hat{\mbq}^{\alpha^*\!(\sigma)}}$ is a $\hat{\mbq}^{\alpha^*\!(\sigma)}$-Brownian motion defined by
\begin{equation}
	B_t^{\hat{\mbq}^{\alpha^*\!(\sigma)}}:=\eta(\sigma)\int_0^t\sigma\sqrt{\nu_s}ds+\hat{B}_t^{\alpha^{*}(\sigma)},\quad 0\leq t\leq T.
\end{equation}
The martingale extraction method shows that
\begin{equation}
	V(T\,;\,\alpha^*(\sigma))=e^{-\lambda(\sigma)T-\eta(\sigma)\nu_0}\mathbb{E}^{\hat{\mbq}^{\alpha^{*}\!(\sigma)}}[e^{\eta(\sigma)\nu_T}].
\end{equation}
Clearly, for any $T\geq 0$ and fixed $\sigma_0\in[\underline{\sigma},\overline{\sigma}]$,
\begin{equation}
	1\leq \inf_{\sigma\in[\underline{\sigma},\overline{\sigma}],T\geq 0}\mathbb{E}^{\hat{\mbq}^{\alpha^{*}\!(\sigma)}}[e^{\eta(\sigma)\nu_T}]\leq \mathbb{E}^{\hat{\mbq}^{\alpha^{*}\!(\sigma_0)}}[e^{\eta(\sigma_0)\nu_T}],
\end{equation}
and $\mathbb{E}^{\hat{\mbq}^{\alpha^{*}\!(\sigma_0)}}[e^{\eta(\sigma_0)\nu_T}]$ converges to some positive constant (refer to Subsection \ref{sec:CIR}). Hence,
\begin{equation}
	\lim_{T\to\infty}\frac{1}{T}\log{\inf_{\sigma\in[\underline{\sigma},\overline{\sigma}]}\mathbb{E}^{\hat{\mbq}^{\alpha^{*}\!(\sigma)}}[e^{\eta(\sigma)\nu_T}]}=0,
\end{equation}
and
\begin{equation} \label{liminf:V:Heston}
	\liminf_{T\to\infty}\frac{1}{T}\log{\inf_{\sigma\in[\underline{\sigma},\overline{\sigma}]}V(T\,;\,\alpha^*(\sigma))}\geq \inf_{\sigma\in[\underline{\sigma},\overline{\sigma}]}-\lambda(\sigma)=-\overline{b}\sup_{\sigma\in[\underline{\sigma},\overline{\sigma}]}\eta(\sigma).
\end{equation}

Conversely, choose $\sigma^*\in[\underline{\sigma},\overline{\sigma}]$ achieving the supremum of $\eta(\sigma)$. Indeed, such $\sigma^*$ exists because the function $\sigma\mapsto\eta(\sigma)$ is continuous. Based on the earlier computations,
\begin{equation} \label{lim:V:Heston}
	\lim_{T\to\infty}\frac{1}{T}\log{V(T\,;\,\alpha^*(\sigma^*))}=-\overline{b}\,\eta(\sigma^*).
\end{equation}
Combining \eqref{liminf:V:Heston} and \eqref{lim:V:Heston} yields
\begin{equation}
	\lim_{T\to\infty}\frac{1}{T}\log{\inf_{\sigma\in[\underline{\sigma},\overline{\sigma}]}V(T\,;\,\alpha^*(\sigma))}=-\overline{b}\,\eta(\sigma^*).
\end{equation}
Finally, in conjunction with \eqref{eqn:V,v:SV}, we have the convergence
\begin{equation}
	\lim_{T\to\infty}\frac{1}{T}\log{v_T}=p(r+\beta(\mu^{*}-r))-\overline{b}\max_{\sigma\in[\underline{\sigma},\overline{\sigma}]}\eta(\sigma).
\end{equation}

\begin{proposition} \label{prop:Heston}
	Let $0<\underline{\mu}\leq\overline{\mu},\,0<\underline{a}\leq\overline{a},\,0<\underline{\sigma}\leq\overline{\sigma},\,\overline{\sigma}^2/2<\underline{b}\leq\overline{b}$ and $X^{\alpha}$ be the Heston model \eqref{eqn:Heston} with set of parameters $\alpha=(\mu,\rho,b,a,\sigma)$ ranging over $[\underline{\mu},\overline{\mu}]\times[\underline{\rho},\overline{\rho}]\times[\underline{b},\overline{b}]\times[\underline{a},\overline{a}]\times[\underline{\sigma},\overline{\sigma}]$.
	Then, the long-term growth rate of the worst-case expected utility of the LETF $L^{\alpha}=(L_t^{\alpha})_{t\geq 0}$ with the reference process $X^{\alpha}$ is given by
	\begin{equation}
		\lim_{T\to\infty}\frac{1}{T}\log{\inf_{\alpha\in[\underline{\alpha},\overline{\alpha}]}\mathbb{E}^{\mbp^{\alpha}}[L_T^p]}=p(r+\beta(\mu^{*}(\beta)-r))-\overline{b}\,\eta(\sigma^*(\beta),\beta),
	\end{equation}
	where
	\begin{equation}
		\eta(\sigma,\beta)=\frac{1}{\sigma^2}\left(\sqrt{\left(\underline{a}-p\beta\rho^*(\beta)\sigma\right)^2+p(1-p)\beta^2\sigma^2}-\left(\underline{a}-p\beta\rho^*(\beta)\sigma\right)\right),
	\end{equation}
	\begin{equation}
		\mu^{*}(\beta)=\begin{cases}
			\underline{\mu}, &  \beta\geq 0 \\ \overline{\mu}, & \beta<0,
		\end{cases},\qquad \rho^*(\beta)=\begin{cases}
			\overline{\rho} & \beta\geq 0 \\ \underline{\rho} & \beta<0
		\end{cases},
	\end{equation}
	and $\sigma^*(\beta)$ maximizes $\eta$ on $[\underline{\sigma},\overline{\sigma}]$ for each $\beta\in[\underline{\beta},\overline{\beta}]$, provided $\underline{a}-p\beta\rho^*(\beta)\overline{\sigma}>0$. The long-run limit is achieved for $(\mu^*(\beta),\rho^*(\beta),\overline{b},\underline{a},\sigma^*(\beta))$.
\end{proposition}

Although the worst-case set of parameters over the uncertainty set clearly exists, determining the optimal leverage ratio $\beta^*$ explicitly is difficult because the process requires finding $\sigma^*$, computing $\eta(\sigma^*(\beta),\beta)$ explicitly for each $\beta$, and comparing the limits for all $\beta\in[\underline{\beta},\overline{\beta}]$, which are nearly impossible to perform by hand. 
Instead, we numerically present the long-term growth rate for each $\beta$ and optimal ratio $\beta^*$ for a specific range of the parameters. First, to find a suitable mesh size for each given error bound, we define \begin{equation}
	\Lambda(\beta):=p(r+\beta(\mu^{*}(\beta)-r))-\overline{b}\,\eta(\sigma^*(\beta),\beta).
\end{equation}
Since
\begin{align}
	\left|\frac{\eta(\sigma^*(\beta+h),\beta+h)-\eta(\sigma^*(\beta),\beta)}{h}\right|&=\left|\frac{\sup_{\sigma}\eta(\sigma,\beta+h)-\sup_{\sigma}\eta(\sigma,\beta)}{h}\right|\\ &\leq\sup_{\sigma}\left|\frac{\eta(\sigma,\beta+h)-\eta(\sigma,\beta)}{h}\right|\\ &=\sup_{\sigma}\sup_{\beta}|\eta_\beta(\sigma,\beta)|,
\end{align}
where $\eta_{\beta}$ denotes the partial derivative of $\eta$ with respect to $\beta$, we have
\begin{align}
	\lvert\Lambda(\beta+h)-\Lambda(\beta)\rvert&=\left|ph(\mu^*(\beta)-r)-\overline{b}\left(\eta(\sigma^*(\beta+h),\beta+h)-\eta(\sigma^*(\beta),\beta)\right)\right|\\
	&\leq  (p(\overline{\mu}-r)+\overline{b}\sup_{\sigma,\beta}|\eta_{\beta}(\sigma,\beta)|)h.
\end{align}
Set $M:=p(\overline{\mu}-r)+\overline{b}\sup_{\sigma,\beta}|\eta_{\beta}(\sigma,\beta)|$. Then, from the aforementioned inequality we shows that for each $\epsilon>0$,
\begin{equation}
	\lvert\Lambda(\beta+h)-\Lambda(\beta)\rvert\leq\epsilon
\end{equation}
whenever $h$ is less than or equal to $\epsilon/M$. In the Heston model, we can show that
\begin{equation}
	\sup_{\sigma,\beta}\left|\eta_{\beta}(\sigma,5)\right|=\sup_{\sigma}\frac{p}{\sigma}\max\left\{\scriptstyle\frac{\underline{a}\underline{\rho}-\sigma\underline{\beta}(1-p(1-\underline{\rho}^2))}{\sqrt{\left(\underline{a}-p\underline{\beta}\underline{\rho}\sigma\right)^2+p(1-p)\underline{\beta}^2\sigma^2}}-\underline{\rho},\,\frac{-\underline{a}\overline{\rho}+\sigma\overline{\beta}(1-p(1-\overline{\rho}^2))}{\sqrt{\left(\underline{a}-p\overline{\beta}\overline{\rho}\sigma\right)^2+p(1-p)\overline{\beta}^2\sigma^2}}+\overline{\rho}\right\}.
\end{equation}
We set 
$\beta\in[-5,5],\,p=0.5,\,r=0.015,\,[\underline{\mu},\overline{\mu}]=[0.05,0.08],\,[\underline{\rho},\overline{\rho}]=[-0.93,-0.75],\,[\underline{b},\overline{b}]=[0.1,0.2],\,[\underline{a},\overline{a}]=[3,10],\,[\underline{\sigma},\overline{\sigma}]=[0.82,0.93]$. The optimal leverage ratio and the corresponding long-term growth rate are approximately 1.25 ad 0.0179, which are within an error range of 0.05 and 0.01, respectively.
\begin{figure}[h] 
	\centering\includegraphics[width=0.8\textwidth, height=0.4\textheight]{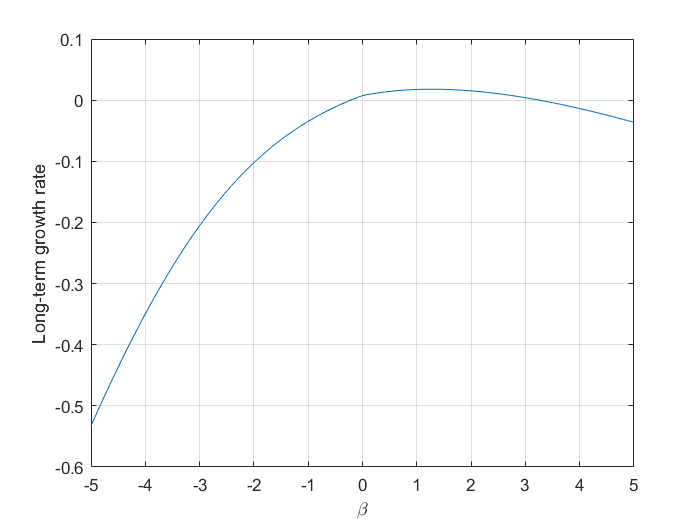}
	\caption{Long-term growth rate of the worst-case expected utility as a function of the leverage ratio $\beta$ under the Heston model.} \label{fig:Heston}
\end{figure}

\newpage

\begin{figure}[h]
	\begin{center} 
		\includegraphics[width=0.45\textwidth, height=0.225\textheight]{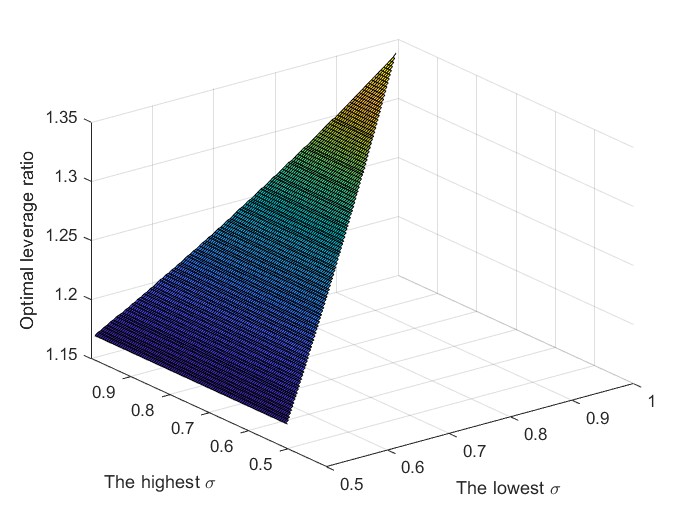} 
		\includegraphics[width=0.45\textwidth, height=0.225\textheight]{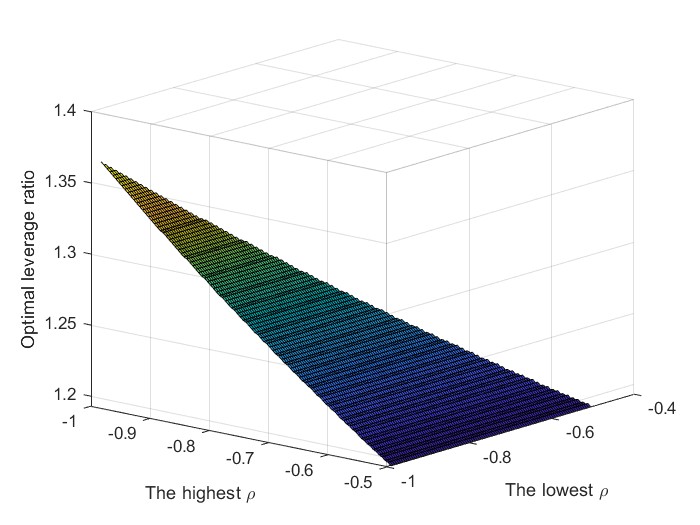}
		\caption{Corresponding optimal leverage ratios to different ranges of $\sigma$ and $\rho$.}\label{fig2:Heston}
	\end{center}
\end{figure}

\begin{figure}[h] 
	\centering\includegraphics[width=0.6\textwidth, height=0.3\textheight]{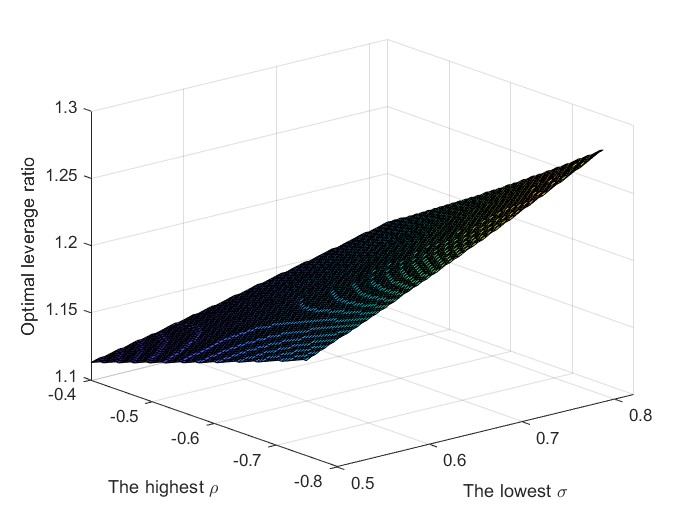}
	\caption{Increase and decrease in the optimal leverage ratio with $\underline{\sigma}$ and $\overline{\rho}$, respectively. } \label{fig3:Heston}
\end{figure}

When the maximum possible boundaries for $\sigma$ and $\rho$ are $[0.5,1]$ and $[-1,-0.5]$ respectively, the optimal leverage ratio increases and decreases in $\underline{\sigma}$ and $\overline{\rho}$, as shown in Figure \ref{fig2:Heston} and \ref{fig3:Heston}, respectively. That is to say, the optimal leverage ratio increases as the lowest possible volatility of the volatility process increases and the highest possible correlation decreases under the Heston model.
\newpage
The obtained results can be extended to the case where $\mu$ is a progressively measurable process taking values in $[\underline{\mu},\overline{\mu}]$ and $b$ and $a$ are the Markovian controls (i.e. there exist functions $b_0:[0,\infty)\times(0,\infty)\to[\underline{b},\overline{b}],\,a_0:[0,\infty)\times(0,\infty)\to[\underline{a},\overline{a}]$ such that $b(t,\omega)=b_0(t,X_t(\omega)),\,a(t,\omega)=a_0(t,X_t(\omega))$) because at least the existence and uniqueness of the SDE
\begin{equation}
	\begin{aligned}
		&dX_t=\mu_tX_t\, dt+ \sqrt{\nu_t}\,X_t\,dW_t^{\rho(t)}\;,\\
		&d\nu_t=(b_0(t,\nu_t)-a_0(t,\nu_t)\nu_t)\,dt+\sigma\sqrt{\nu_t}\,dB_t^{\rho(t)}\;,
	\end{aligned}
\end{equation}
in the weak sense are guaranteed \cite[Chapter 10]{stroock1997}, and the comparison principle remains valid.

\begin{cor}
	Let the reference process $X$ follows
	\begin{equation} \label{Heston:extended}
		\begin{aligned}
			&dX_t=\mu_tX_t\, dt+ \sqrt{\nu_t}\,X_t\,dW_t^{\rho(t)}\;,\\
			&d\nu_t=(b_0(t,\nu_t)-a_0(t,\nu_t)\nu_t)\,dt+\sigma\sqrt{\nu_t}\,dB_t^{\rho(t)}\;,
		\end{aligned}
	\end{equation}
	where $\mu:\Omega\times[0,\infty)\to[\underline{\mu},\overline{\mu}]$ is progressively measurable and $b_0,\,a_0$, and $\rho$ range over $[\underline{b},\overline{b}],\,[\underline{a},\overline{a}]$, and $[\underline{\rho},\overline{\rho}]$, respectively. Then, Proposition \ref{prop:Heston} holds for the LETF with the reference process \eqref{Heston:extended}.
\end{cor}
\begin{remark}
	Furthermore, $b$ and $a$ can be merely progressively measurable processes whenever the existence and uniqueness of the SDE
	\begin{equation}
		\begin{aligned}
			&dX_t=\mu_tX_t\, dt+ \sqrt{\nu_t}\,X_t\,dW_t^{\rho(t)}\;,\\
			&d\nu_t=(b_t-a_t\nu_t)\,dt+\sigma\sqrt{\nu_t}\,dB_t^{\rho(t)}\;,
		\end{aligned}
	\end{equation} are guaranteed.
\end{remark}

\subsection{3/2 volatility model}
Several studies assert that the 3/2 volatility model outperforms the Heston model in that the 3/2 model better captures the volatility smiles (\citet{drimus2012options}) and evolution of the volatility index (\citet{goard2013stochastic}). Unlike the Heston model, the volatility process of the 3/2 model follows the 3/2 process, such that
\begin{equation}\label{eqn:3/2_vol}
	\begin{aligned}
		&dX_t=\mu X_t\, dt+ \sqrt{\nu_t}X_t\,dW_t\;,\\
		&d\nu_t=(b-a \nu_t)\nu_t\,dt+\sigma \nu_t^{3/2}\,dB_t\;,
	\end{aligned}
\end{equation}
where $W_t$ and $B_t$ are two standard Brownian motions with instantaneous correlation $\rho \in[\underline{\rho},\overline{\rho}]$. Additionally, the model has uncertainties in $\alpha=(\mu,\rho,b,a,\sigma)\in[\underline{\alpha},\overline{\alpha}]=[\underline{\mu},\overline{\mu}]\times[\underline{\rho},\overline{\rho}]\times[\underline{b},\overline{b}]\times[\underline{a},\overline{a}]\times[\underline{\sigma},\overline{\sigma}]$ with $\underline{\mu},\underline{b},\underline{\sigma}>0$, and $\underline{a}>-\underline{\sigma}^2/2$.

Further, assuming that $\underline{a}-p|\beta|\overline{\sigma}>-\underline{\sigma}^2/2$ so that $a-p\beta\rho\,\sigma>-\underline{\sigma}^2/2$ for every $(a,\rho,\sigma)\in[\underline{a},\overline{a}]\times[\underline{\rho},\overline{\rho}]\times[\underline{\sigma},\overline{\sigma}]$, the SDE
\begin{equation}
	d\nu_t=(b-(a-p\beta\rho\,\sigma)\nu_t)\nu_tdt+\sigma\nu_t^{3/2}\,dB_t,\quad \nu_0>0,
\end{equation}
has a unique strong solution. Hence, a probability measure $\hat{\mbp}^{\alpha}$ defined on $\mathcal{F}_T$ for each $T\geq0$ by \eqref{measure:SV} makes sense, and $\nu$ follows
\begin{equation}
	d\nu_t=(b-(a-p\beta\rho\,\sigma)\nu_t)\nu_tdt+\sigma\nu_t^{3/2}\,d\hat{B}_t^{\alpha}
\end{equation}
under $\hat{\mbp}^{\alpha}$, with a $\hat{\mbp}^{\alpha}$-Brownian motion
$$\hat{B}_t^{\alpha}=-p\beta\rho\int_0^t\sqrt{\nu_s}\,ds+B_t.$$
Moreover, by applying the comparison principle to $\nu$, we have
\begin{equation}
	\mathbb{E}^{\mbp^{\alpha}}[L_T^p]\geq e^{p(r+\beta(\mu^{*}-r))T}\mathbb{E}^{\hat{\mbp}^{\mu^{*},\rho^*,\overline{b},\underline{a},\sigma}}[e^{-\frac{1}{2}p(1-p)\beta^2\int_0^T\nu_sds}],\quad T\geq 0,
\end{equation}
for every $\alpha=(\mu,\rho,b,a,\sigma)$, where
\begin{equation}
	\rho^*=\begin{cases}
		\overline{\rho} & \beta\geq 0 \\ \underline{\rho} & \beta<0
	\end{cases},
\end{equation}
To simplify the notation, set
\begin{equation}
	\alpha^{*}(\sigma):=(\mu^{*},\rho^*,\overline{b},\underline{a},\sigma).
\end{equation}

Under the measure $\hat{\mbp}^{\alpha^{*}(\sigma)}$, the eigenpair problem for the infinitesimal generator of $\nu$ is expressed as
$$-\lambda\phi(\nu)=\frac{1}{2}\sigma^2\nu^3\phi''(\nu)+(\overline{b}-(\underline{a}-p\beta\rho^*\sigma) \nu)\nu\,\phi'(\nu)-\frac{1}{2}p(1-p)\beta^2\nu\,\phi(\nu)\;.$$
One solution pair is given by
\begin{equation*}
	(\lambda(\sigma),\phi_{\sigma}(v))=\left(\overline{b}\,\eta(\sigma),\nu^{-\eta(\sigma)}\right) \,,
\end{equation*}
where
\begin{equation}
	\eta(\sigma)=\frac{1}{\sigma^2}\left(\sqrt{\left(\underline{a}-p\beta\rho^*\sigma+\sigma^2/2\right)^2+p(1-p)\beta^2\sigma^2}-\left(\underline{a}-p\beta\rho^*\sigma+\sigma^2/2\right)\right).
\end{equation}
As aforementioned, a local martingale
$$M_t:=\exp{\left\{\lambda(\sigma) t-\frac{1}{2}p\beta(\beta-1)\int_0^t\nu_sds-\eta(\sigma)\nu_t+\eta(\sigma)\nu_0\right\}},\quad 0\leq t \leq T$$
can be shown as a true martingale; hence, for a probability measure $\hat{\mbq}^{\alpha^*\!(\sigma)}$ defined by 
$$\frac{d\hat{\mathbb{Q}}^{\alpha^{*}\!(\sigma)}}{d\hat{\mathbb{P}}^{\alpha^{*}\!(\sigma)}}\biggl|_{\mathcal{F}_T}=M_T$$
on $\mathcal{F}_T$, the process $\nu$ under $\hat{\mbq}^{\alpha^*\!(\sigma)}$ satisfies
\begin{equation}
	d\nu_t=\left(\overline{b}-\left(\sqrt{(\underline{a}-p\beta\rho^*\sigma+\sigma^2/2)^2+p(1-p)\beta^2\sigma^2}-\sigma^2/2\right)\nu_t\right)\nu_t\,dt+\sigma\,\nu_t^{3/2}dB_t^{\alpha^*\!(\sigma)},
\end{equation}
where $B^{\hat{\mbq}^{\alpha^*\!(\sigma)}}$ is a $\hat{\mbq}^{\alpha^*\!(\sigma)}$-Brownian motion defined by
\begin{equation}
	B_t^{\hat{\mbq}^{\alpha^*\!(\sigma)}}:=\eta(\sigma)\int_0^t\sigma\sqrt{\nu_s}ds+\hat{B}_t^{\alpha^{*}(\sigma)}.
\end{equation}
Thus,
\begin{equation}
	V(T\,;\,\alpha^*(\sigma))=e^{-\lambda(\sigma)T}\nu_0^{-\eta(\sigma)}\mathbb{E}^{\hat{\mbq}^{\alpha^{*}\!(\sigma)}}[\nu_T^{\eta(\sigma)}].
\end{equation}
We show that
\begin{equation}
	\lim_{T\to\infty}\frac{1}{T}\log{\inf_{\sigma\in[\underline{\sigma},\overline{\sigma}]}\mathbb{E}^{\hat{\mbq}^{\alpha^{*}\!(\sigma)}}[\nu_T^{\eta(\sigma)}]}=0,
\end{equation}
leading to
\begin{equation} \label{liminf:V:3/2_vol}
	\liminf_{T\to\infty}\frac{1}{T}\log{\inf_{\sigma\in[\underline{\sigma},\overline{\sigma}]}V(T\,;\,\alpha^*(\sigma))}\geq \inf_{\sigma\in[\underline{\sigma},\overline{\sigma}]}-\lambda(\sigma)=-\overline{b}\sup_{\sigma\in[\underline{\sigma},\overline{\sigma}]}\eta(\sigma).
\end{equation}
To this end, we claim that for any fixed $\sigma_0\in[\underline{\sigma},\overline{\sigma}]$ and $T\geq 0$,
\begin{equation}
	0< \inf_{\sigma\in[\underline{\sigma},\overline{\sigma}],T\geq 0}\mathbb{E}^{\hat{\mbq}^{\alpha^{*}\!(\sigma)}}[\nu_T^{\eta(\sigma)}]\leq \mathbb{E}^{\hat{\mbq}^{\alpha^{*}\!(\sigma_0)}}[\nu_T^{\eta(\sigma_0)}].
\end{equation}
The second inequality is trivial, and $\mathbb{E}^{\hat{\mbq}^{\alpha^{*}\!(\sigma_0)}}[e^{\eta(\sigma_0)\nu_T}]$ converges to some positive constant as $T\to\infty$ (refer to Subsetion \ref{sec:3/2_model}). For the first inequality, for each $\sigma\in[\underline{\sigma},\overline{\sigma}]$ and $T\geq 0$, the expectation $\mathbb{E}^{\hat{\mbq}^{\alpha^{*}\!(\sigma)}}[\nu_T^{\eta(\sigma)}]$ is positive. In fact, $\inf_{T\geq 0}\mathbb{E}^{\hat{\mbq}^{\alpha^{*}\!(\sigma)}}[\nu_T^{\eta(\sigma)}]>0$ because the function $T\mapsto\mathbb{E}^{\hat{\mbq}^{\alpha^{*}\!(\sigma)}}[\nu_T^{\eta(\sigma)}]$ is continuous on $[0,\infty)$ \cite[Lemma D.2]{park2023dynamic} and converges to some positive constant as $T\to\infty$. Thus, $\inf_{\sigma\in[\underline{\sigma},\overline{\sigma}]}\inf_{T\geq 0}\mathbb{E}^{\hat{\mbq}^{\alpha^{*}\!(\sigma)}}[\nu_T^{\eta(\sigma)}]>0$ since the function $\sigma\mapsto\inf_{T\geq 0}\mathbb{E}^{\hat{\mbq}^{\alpha^{*}\!(\sigma)}}[\nu_T^{\eta(\sigma)}]$ is continuous on $[\underline{\sigma},\overline{\sigma}]$.

Conversely, since the function $\sigma\mapsto\eta(\sigma)$ is continuous, there exists $\sigma^*\in[\underline{\sigma},\overline{\sigma}]$ achieving the supremum of $\eta(\sigma)$. Applying the aforementioned calculations to $V(T\,;\,\alpha^*(\sigma^*))$, we show that
\begin{equation} \label{lim:V:3/2_vol}
	\lim_{T\to\infty}\frac{1}{T}\log{V(T\,;\,\alpha^*(\sigma^*))}=-\overline{b}\,\eta(\sigma^*).
\end{equation}
Combining \eqref{liminf:V:3/2_vol} and \eqref{lim:V:3/2_vol} yields
\begin{equation}
	\lim_{T\to\infty}\frac{1}{T}\log{\inf_{\sigma\in[\underline{\sigma},\overline{\sigma}]}V(T\,;\,\alpha^*(\sigma))}=-\overline{b}\,\eta(\sigma^*).
\end{equation}
Therefore, the convergence
\begin{equation}
	\lim_{T\to\infty}\frac{1}{T}\log{v_T}=p(r+\beta(\mu^{*}-r))-\overline{b}\max_{\sigma\in[\underline{\sigma},\overline{\sigma}]}\eta(\sigma).
\end{equation}
is derived from \eqref{eqn:V,v:SV}.

\begin{proposition} \label{prop:3/2_vol}
	Let $0<\underline{\mu}\leq\overline{\mu},\,0<\underline{b}\leq\overline{b},\,0<\underline{\sigma}\leq\overline{\sigma},\,-\underline{\sigma}^2/2<\underline{a}\leq\overline{a}$ and $X^{\alpha}$ be the 3/2 volatility model \eqref{eqn:3/2_vol} with set of parameters $\alpha=(\mu,\rho,b,a,\sigma)$ ranging over $[\underline{\alpha},\overline{\alpha}]=[\underline{\mu},\overline{\mu}]\times[\underline{\rho},\overline{\rho}]\times[\underline{b},\overline{b}]\times[\underline{a},\overline{a}]\times[\underline{\sigma},\overline{\sigma}]$.
	Then, the long-term growth rate of the worst-case expected utility of the LETF $L^{\alpha}=(L_t^{\alpha})_{t\geq 0}$, with the reference process $X^{\alpha}$ is given by
	\begin{equation}
		\lim_{T\to\infty}\frac{1}{T}\log{\inf_{\alpha\in[\underline{\alpha},\overline{\alpha}]}\mathbb{E}^{\mbp^{\alpha}}[L_T^p]}=p(r+\beta(\mu^{*}(\beta)-r))-\overline{b}\,\eta(\sigma^*(\beta),\beta),
	\end{equation}
	where
	\begin{equation}
		\eta(\sigma,\beta)=\frac{\sqrt{\left(\underline{a}-p\beta\rho^*(\beta)\sigma+\sigma^2/2\right)^2+p(1-p)\beta^2\sigma^2}-\left(\underline{a}-p\beta\rho^*(\beta)\sigma+\sigma^2/2\right)}{\sigma^2},
	\end{equation}
	\begin{equation}
		\mu^{*}(\beta)=\begin{cases}
			\underline{\mu}, &  \beta\geq 0 \\ \overline{\mu}, & \beta<0,
		\end{cases},\qquad \rho^*(\beta)=\begin{cases}
			\overline{\rho} & \beta\geq 0 \\ \underline{\rho} & \beta<0
		\end{cases},
	\end{equation}
	and $\sigma^*(\beta)$ maximizes $\eta$ on $[\underline{\sigma},\overline{\sigma}]$ for each $\beta\in[\underline{\beta},\overline{\beta}]$, provided $\underline{a}-p\beta\rho^*(\beta)\overline{\sigma}+\underline{\sigma}^2/2>0$. The long-run limit is achieved for $\alpha=(\mu^*(\beta),\rho^*(\beta),\overline{b},\underline{a},\sigma^*(\beta))$.
\end{proposition}

The obtained results are similar to those of the Heston model, indicating that the difficulty of computing the optimal leverage ratio $\beta^*$ by hand. Instead, the numerical approach of finding the optimal leverage ratio and corresponding robust long-term growth rate adopted in the aforementioned subsection works for this model as well.

Generalization of coefficients of the 3/2 volatility process is discussed in the next corollary.
\begin{cor}
	Let the reference process $X$ follow
	\begin{equation}
		\begin{aligned}
			&dX_t=\mu_tX_t\, dt+ \sqrt{\nu_t}\,X_t\,dW_t^{\rho(t)}\;,\\
			&d\nu_t=(b_0(t,\nu_t)-a_0(t,\nu_t)\nu_t)\nu_t\,dt+\sigma\nu_t^{3/2}\,dB_t^{\rho(t)}\;,
		\end{aligned}
	\end{equation}
	where $\mu$ is a progressively measurable process taking values in $[\underline{\mu},\overline{\mu}]$, and $b_0,\,a_0$ and $\rho$ are functions mapping to $[\underline{b},\overline{b}],\,[\underline{a},\overline{a}]$, and $[\underline{\rho},\overline{\rho}]$, respectively. Then, Proposition \ref{prop:3/2_vol} holds for the LETF with the reference $X$.
\end{cor}

\section{Uncertainties on reference and interest rate}
\label{sec:interest_rate}
In this section, we consider the short interest rate as stochastic. Precisely, we deal with the models in the form
\begin{align}
	&dX_t=\mu X_t\,dt+\varsigma X_t\,dW_t\;,\\
	&dr_t=b(r_t)\,dt+\sigma(r_t)\,dB_t\;,\quad r_0>0,
\end{align}
where $W_t$ and $B_t$ are two Brownian motions such that $\langle W,B\rangle_t=\rho\,t$ with  $-1\leq\underline{\rho}\leq\rho\leq\overline{\rho}\leq 1$, and $r_0$ is deterministic. We assume that uncertainties lie in $\mu,\,\varsigma,\,\rho$, and the set of parameters $\tilde{\alpha}$ of $r$. Thus, $\alpha=(\mu,\varsigma,\rho,\tilde{\alpha})$ and $[\underline{\alpha},\overline{\alpha}]=[\underline{\mu},\overline{\mu}]\times[\underline{\varsigma},\overline{\varsigma}]\times[\underline{\rho},\overline{\rho}]\times[\underline{\tilde{\alpha}},\overline{\tilde{\alpha}}]$ with $\underline{\mu},\underline{\varsigma}>0$. This section also introduce the expression
\begin{align}
	&dX_t^{\alpha}=\mu X_t^{\alpha}\,dt+\varsigma X_t^{\alpha}\,dW_t^{\rho}\;,\\
	&dr_t^{\alpha}=b(r_t^{\alpha};\alpha)\,dt+\sigma(r_t^{\alpha};\alpha)\,dB_t^{\rho}\;,
\end{align}
as in Section \ref{sec:SV}. Then, the $p$-th power of the corresponding LETF value and expected utility of an investor at time $T\geq 0$ are given by
\begin{align}
	L_T^p&=e^{p(\beta\mu-\beta^2\varsigma^2/2)T-p(\beta-1)\int_0^Tr_s^{\alpha}ds+p\beta\varsigma W_T^{\rho}}\\
	&=e^{p\beta\mu T-p(1-p)\beta^2\varsigma^2T/2-p(\beta-1)\int_0^Tr_s^{\alpha}ds}\mathcal{E}\!\left(p\beta\varsigma W_{\cdot}^{\rho}\right)_T,
\end{align}
and
\begin{equation} \label{utility:interest_rate}
	\mathbb{E}^{\mathbb{P}^{\alpha}}[L_T^p]=\mathbb{E}^{\hat{\mbp}^{\alpha}}[e^{p\beta\mu T-p(1-p)\beta^2\varsigma^2T/2-p(\beta-1)\int_0^Tr_s^{\alpha}ds}],\quad T\geq 0,
\end{equation}
respectively, where the probability measure $\hat{\mathbb{P}}^{\alpha}$ is defined on $\mathcal{F}_T$ for each $T\geq 0$ by
\begin{equation} \label{RN:StocIR}
	\frac{d\hat{\mathbb{P}}^{\alpha}}{d\mbp}\biggl|_{\mathcal{F}_T}=\mathcal{E}\!\left(p\beta\varsigma W_{\cdot}^{\rho}\right)_T,
\end{equation}
under which the $\hat{\mathbb{P}}^{\alpha}$-dynamics of $(X^{\alpha},r^{\alpha})$ satisfies
\begin{equation}
	\begin{aligned}
		dX_t^{\alpha}&=(\mu+p\beta\varsigma^2)X_t^{\alpha}\,dt+\varsigma X_t^{\alpha}\,d\hat{W}_t,\\
		dr_t^{\alpha}&=\left(b(r_t^{\alpha};\alpha)+p\beta\varsigma\rho\sigma(r_t^{\alpha};\alpha)\right)\,dt+\sigma(r_t^{\alpha};\alpha)\,d\hat{B}_t^{\alpha},\quad 0\leq t\leq T,
	\end{aligned}
\end{equation}
for two standard $\hat{\mathbb{P}}^{\alpha}$-Brownian motion
\begin{align}
	\hat{W}_t^{\alpha}&=-p\beta\varsigma\,t+W_t,\\
	\hat{B}_t^{\alpha}&=-p\beta\varsigma\rho\,t+B_t,\quad 0\leq t\leq T.
\end{align}
For each $\alpha\in[\underline{\alpha},\overline{\alpha}]$, we have
\begin{equation} \label{temp:interest_rate}
	\mathbb{E}^{\mathbb{P}^{\alpha}}[L_T^p]\geq e^{p\beta\mu^{*} T}\mathbb{E}^{\hat{\mbp}^{\alpha}}[e^{-p(1-p)\beta^2\varsigma^2T/2-p(\beta-1)\int_0^Tr_s^{\alpha}ds}],\quad T\geq 0,
\end{equation}
from \eqref{utility:interest_rate}, where
\begin{equation} \label{mu*}
	\mu^{*}=\begin{cases}
		\underline{\mu}, &  \beta\geq 0 \\ \overline{\mu}, & \beta<0.
	\end{cases}
\end{equation}
The roles of $V$ and $v$ in this section are
\begin{align}
	V(T\,;\,\alpha)&=\mathbb{E}^{\hat{\mbp}^{\alpha}}[e^{-p(1-p)\beta^2\varsigma^2T/2-p(\beta-1)\int_0^Tr_s^{\alpha}ds}],\\
	v_T&=\inf_{\alpha\in[\underline{\alpha},\overline{\alpha}]}\EP[(L_T^{\alpha})^p]=e^{p\beta\mu^{*} T}\inf_{\alpha\in[\underline{\alpha},\overline{\alpha}]}V(T\,;\,\alpha). \label{eqn:V,v:interest_rate}
\end{align}
Thus, the comparison principle for $r^{\alpha}$ and the martingale extraction method applied to $V(T\,;\,\alpha)$ serve as the main tools.

\subsection{Vasicek interest rate}
\label{sec:Vasicek}
The Vasicek interest rate model was named after \citet{vasicek1977equilibrium}.
The model considers GBM as the reference and Ornstein--Uhlenbeck process as the short interest rate, such that
\begin{equation} \label{SDE:vasicek}
	\begin{aligned}
		dX_t&=\mu X_t\,dt+\varsigma X_t\,dW_t,\\
		dr_t&=(b-a r_t)\,dt+\sigma \,dB_t,
	\end{aligned}
\end{equation}
where $W$ and $B$ are two Brownian motions, $\langle W,B\rangle_t=\rho\,t$ with  $-1\leq\underline{\rho}\leq\rho\leq\overline{\rho}\leq 1$. This model has uncertainties in $\alpha=(\mu,\varsigma,\rho,b,a,\sigma)\in[\underline{\alpha},\overline{\alpha}]=[\underline{\mu},\overline{\mu}]\times[\underline{\varsigma},\overline{\varsigma}]\times[\underline{\rho},\overline{\rho}]\times[\underline{b},\overline{b}]\times[\underline{a},\overline{a}]\times[\underline{\sigma},\overline{\sigma}]$ with $\underline{\mu},\underline{\varsigma},\underline{b},\underline{a},\underline{\sigma}>0$. Under the probability measure $\hat{\mbp}^{\alpha}$ defined by \eqref{RN:StocIR}, the process \eqref{SDE:vasicek} follows
\begin{equation}
	\begin{aligned}
		dX_t&=(\mu+p\beta\varsigma^2)X_t\,dt+\varsigma X_t\,d\hat{W}_t,\\
		dr_t&=\left(b+p\beta\varsigma\rho\sigma-ar_t\right)\,dt+\sigma\,d\hat{B}_t^{\alpha},\quad 0\leq t\leq T.
	\end{aligned}
\end{equation}
Here, the interest rate can take negative values, implying that the comparison principle cannot determine the parameter $a$ when it comes to finding parameters achieving the worst-case scenario.  Moreover, we will see that the parameters $\varsigma,\,a,$ and $\sigma$ are determined not only by the sign of $\beta$ but also by the signs of $\underline{\rho}$ and $\overline{\rho}$.

The eigenpair problem for the infinitesimal generator of $r$ is expressed as
$$-\lambda\phi(r)=\frac{1}{2}\sigma^2\phi''(r)+(b+p\beta\varsigma\rho\sigma-a\,r)\,\phi'(r)-p(\beta-1)r\,\phi(r),$$
whose one solution pair is given by
\begin{equation*}
	\left(\lambda(\alpha),\phi(r)\right):=\left(-\frac{1}{2}\left(p(\beta-1)\frac{\sigma}{a}\right)^2+p^2\beta(\beta-1)\varsigma\rho\frac{\sigma}{a}+\frac{b\,p(\beta-1)}{a},e^{-\frac{p(\beta-1)r}{a} }\right).
\end{equation*}
Under a probability measure $\hat{\mbq}^{\alpha}$ defined by 
$$\frac{d\hat{\mathbb{Q}}^{\alpha}}{d\hat{\mathbb{P}}^{\alpha}}\biggl|_{\mathcal{F}_T}=\exp{\left\{\lambda(\alpha) T-p(\beta-1)\int_0^T r_s\,ds-\frac{p(\beta-1)}{a}r_T+\frac{p(\beta-1)}{a}r_0  \right\}}$$
on $\mathcal{F}_T$ for each $T>0$, the process $r$ satisfies
$$dr_t=\left(b+p\beta\varsigma\rho\sigma-\frac{p(\beta-1)\sigma^2}{a} -a\,r_t\right)\,dt+\sigma \,dB_t^{\hat{\mbq}^{\alpha}}\;,$$
with a $\hat{\mbq}$-Brownian motion
\begin{equation}
	B_t^{\hat{\mbq}^{\alpha}}=\frac{p(\beta-1)\sigma}{a}t+\hat{B}_t^{\alpha}.
\end{equation}
Thus,
\begin{equation}
	V(T\,;\,\alpha)=e^{-p(1-p)\beta^2\varsigma^2T/2-\lambda(\alpha) T+\frac{p(1-\beta)}{a}r_0}\mathbb{E}^{\hat{\mbq}^{\alpha}}[e^{\frac{p(\beta-1)}{a}r_T}].
\end{equation}
Note that $\mathbb{E}^{\hat{\mbq}^{\alpha}}[e^{\frac{p(\beta-1)}{a}r_T}]$ is the value of the moment generating function of $r_T$ evaluated at $p(\beta-1)/a$. It is well-known that $$r_T\sim\mathcal{N}\left(e^{-a\,T}+\frac{1}{a}\left(b+p\beta\varsigma\rho\sigma-\frac{p(\beta-1)\sigma^2}{a}\right)(1-e^{-a\,T}),\,\frac{\sigma^2}{2a}(1-e^{-2a\,T})\right),$$
with its mean and variance bounded and converging to some positive constant as $T\to\infty$, regardless of the $\alpha$ value. Thus, we readily observe that
\begin{equation}
	\lim_{T\to\infty}\frac{1}{T}\log\inf_{\alpha\in[\underline{\alpha},\overline{\alpha}]}V(T\,;\,\alpha)=\inf_{\alpha\in[\underline{\alpha},\overline{\alpha}]}-\frac{1}{2}p(1-p)\beta^2\varsigma^2-\lambda(\alpha).
\end{equation}
The remaining task involves finding $\alpha^*\in[\underline{\alpha},\overline{\alpha}]$ achieving the infimum.

As aforementioned, the signs of $\underline{\rho}$ and $\overline{\rho}$ affect the value of $\alpha^*$. We denote $\alpha^*=(\mu^*,\varsigma^*,\rho^*,b^*,a^*,\sigma^*)$. Recall that $\mu^*$ is already determined in \eqref{mu*}.

\begin{mycase}
	\case $\beta\in[1,\overline{\beta}]$ and $\overline{\rho}>0$: $b^*=\overline{b},\,\varsigma^*=\overline{\varsigma},\,\rho^*=\overline{\rho}$, and $$ (a^*,\sigma^*)\in\! \argmax_{(a,\sigma)\in[\underline{a},\overline{a}]\times[\underline{\sigma},\overline{\sigma}]}\!\lambda(\overline{\varsigma},\overline{\rho},\overline{b},a,\sigma)=\!\argmax_{(a,\sigma)\in[\underline{a},\overline{a}]\times[\underline{\sigma},\overline{\sigma}]}\!-\frac{1}{2}\left(p(\beta-1)\frac{\sigma}{a}\right)^2\!+p^2\beta(\beta-1)\overline{\varsigma}\overline{\rho}\frac{\sigma}{a}+\frac{p(\beta-1)\overline{b}}{a}.$$
	
	\case $\beta\in[1,\overline{\beta}]$ and $\overline{\rho}<0$: $b^*=\overline{b},\,\rho^*=\overline{\rho},\,\sigma^*=\underline{\sigma}$, and
	\begin{align}
		(\varsigma^*,a^*)\in&\argmax_{(\varsigma,a)\in[\underline{\varsigma},\overline{\varsigma}]\times[\underline{a},\overline{a}]}\frac{1}{2}p(1-p)\beta^2\varsigma^2+\lambda(\varsigma,\overline{\rho},\overline{b},a,\underline{\sigma})\\ &\quad=\argmax_{(\varsigma,a)\in[\underline{\varsigma},\overline{\varsigma}]\times[\underline{a},\overline{a}]}\frac{1}{2}p(1-p)\beta^2\varsigma^2-\frac{1}{2}p^2(\beta-1)^2\frac{\underline{\sigma}^2}{a^2}+p^2\beta(\beta-1)\overline{\rho}\underline{\sigma}\frac{\varsigma}{a}+\frac{p(\beta-1)\overline{b}}{a}.
	\end{align}
	
	\case $\beta\in[0,1)$ and $\underline{\rho}<0$: $b^*=\underline{b},\,\varsigma^*=\overline{\varsigma},\,\rho^*=\underline{\rho}$, and 
	$$ (a^*,\sigma^*)\in\! \argmax_{(a,\sigma)\in[\underline{a},\overline{a}]\times[\underline{\sigma},\overline{\sigma}]}\!\lambda(\overline{\varsigma},\underline{\rho},\underline{b},a,\sigma)=\!\argmax_{(a,\sigma)\in[\underline{a},\overline{a}]\times[\underline{\sigma},\overline{\sigma}]}\!-\frac{1}{2}\left(p(\beta-1)\frac{\sigma}{a}\right)^2\!+p^2\beta(\beta-1)\overline{\varsigma}\underline{\rho}\frac{\sigma}{a}+\frac{p(\beta-1)\underline{b}}{a}.$$
	
	\case $\beta\in[0,1)$ and $\underline{\rho}>0$: $b^*=\underline{b},\,\rho^*=\underline{\rho},\,a^*=\overline{a},\,\sigma^*=\underline{\sigma}$, and
	\begin{equation}
		\varsigma^*=\argmax_{\varsigma\in[\underline{\varsigma},\overline{\varsigma}]}\frac{1}{2}p(1-p)\beta^2\varsigma^2+p^2\beta(\beta-1)\underline{\rho}\frac{\underline{\sigma}}{\overline{a}}\varsigma.
	\end{equation}
	
	\case $\beta\in[\underline{\beta},0)$ and $\overline{\rho}>0$: $b^*=\underline{b},\,\varsigma^*=\overline{\varsigma},\,\rho^*=\overline{\rho}$, and $$ (a^*,\sigma^*)\in\! \argmax_{(a,\sigma)\in[\underline{a},\overline{a}]\times[\underline{\sigma},\overline{\sigma}]}\!\lambda(\overline{\varsigma},\overline{\rho},\underline{b},a,\sigma)=\!\argmax_{(a,\sigma)\in[\underline{a},\overline{a}]\times[\underline{\sigma},\overline{\sigma}]}\!-\frac{1}{2}\left(p(\beta-1)\frac{\sigma}{a}\right)^2\!+p^2\beta(\beta-1)\overline{\varsigma}\overline{\rho}\frac{\sigma}{a}+\frac{p(\beta-1)\underline{b}}{a}.$$
	
	\case $\beta\in[\underline{\beta},0)$ and $\overline{\rho}<0$:  $b^*=\underline{b},\,\rho^*=\overline{\rho},\,a^*=\overline{a},\,\sigma^*=\underline{\sigma}$, and
	\begin{equation}
		\varsigma^*=\argmax_{\varsigma\in[\underline{\varsigma},\overline{\varsigma}]}\frac{1}{2}p(1-p)\beta^2\varsigma^2+p^2\beta(\beta-1)\overline{\rho}\frac{\underline{\sigma}}{\overline{a}}\varsigma.
	\end{equation}
\end{mycase}

\begin{proposition} \label{prop:vasicek}
	Let $0<\underline{\mu}\leq\overline{\mu},\,0<\underline{\varsigma}\leq\overline{\varsigma},\,0<\underline{b}\leq\overline{b},\,0<\underline{a}\leq\overline{a},\,0<\underline{\sigma}\leq\overline{\sigma}$, and $(X^{\alpha},r^{\alpha})$ be the process \eqref{SDE:vasicek} with set of parameters $\alpha=(\mu,\varsigma,\rho,b,a,\sigma)$ ranging over $[\underline{\mu},\overline{\mu}]\times[\underline{\varsigma},\overline{\varsigma}]\times[\underline{\rho},\overline{\rho}]\times[\underline{b},\overline{b}]\times[\underline{a},\overline{a}]\times[\underline{\sigma},\overline{\sigma}]$.
	Then, the long-term growth rate of the worst-case expected utility of the LETF $L^{\alpha}=(L_t^{\alpha})_{t\geq 0}$ with reference process and interest rate $X^{\alpha}$ and $r^{\alpha}$, respectively, is given by
	\begin{equation}
		\lim_{T\to\infty}\frac{1}{T}\log{\inf_{\alpha\in[\underline{\alpha},\overline{\alpha}]}\mathbb{E}^{\mbp^{\alpha}}[L_T^p]}=p\beta\mu^{*}(\beta)-\frac{1}{2}p(1-p)\beta^2\varsigma^*(\beta,\underline{\rho},\overline{\rho})^2-\lambda(\alpha^*(\beta,\underline{\rho},\overline{\rho})),
	\end{equation}
	where
	\begin{align}
		\alpha^*(\beta,\underline{\rho},\overline{\rho})&=\left(\mu^*(\beta),\varsigma^*(\beta,\underline{\rho},\overline{\rho}),\rho^*(\beta),b^*(\beta),a^*(\beta,\underline{\rho},\overline{\rho}),\sigma^*(\beta,\underline{\rho},\overline{\rho})\right)\\
		\lambda(\alpha,\beta)&=-\frac{1}{2}\left(p(\beta-1)\frac{\sigma}{a}\right)^2\!+p^2\beta(\beta-1)\varsigma\rho\frac{\sigma}{a}+\frac{p(\beta-1)b}{a},
	\end{align}
	\begin{equation}
		\mu^{*}(\beta)=\begin{cases}\underline{\mu}, &  \beta\geq 0 \\ \overline{\mu}, & \beta<0
		\end{cases},\,
		\rho^*(\beta)=\begin{cases}
			\overline{\rho}, & \beta\in[\underline{\beta},0)\cup[1,\overline{\beta}] \\ \underline{\rho}, & \beta\in[0,1) 
		\end{cases},\,b^*(\beta)=\begin{cases}\overline{b}, &  \beta\geq 1 \\ \underline{b}, & \beta<1\end{cases},
	\end{equation}
	\begin{equation}
		\begin{pmatrix}
			\varsigma^*(\beta,\underline{\rho},\overline{\rho}) \\ a^*(\beta,\underline{\rho},\overline{\rho}) \\ \sigma^*(\beta,\underline{\rho},\overline{\rho})
		\end{pmatrix}'=\begin{cases}
			(\overline{\varsigma},\argmax\limits_{(a,\sigma)\in[\underline{a},\overline{a}]\times[\underline{\sigma},\overline{\sigma}]}\!\lambda(\overline{\varsigma},\overline{\rho},\overline{b},a,\sigma)) & \beta\in[1,\overline{\beta}],\,\overline{\rho}>0 \\
			(\argmax\limits_{(\varsigma,a)\in[\underline{\varsigma},\overline{\varsigma}]\times[\underline{a},\overline{a}]}\frac{1}{2}p(1-p)\beta^2\varsigma^2+\lambda(\varsigma,\overline{\rho},\overline{b},a,\underline{\sigma}),\underline{\sigma}) & \beta\in[1,\overline{\beta}],\,\overline{\rho}<0 \\ 
			(\overline{\varsigma},\argmax\limits_{(a,\sigma)\in[\underline{a},\overline{a}]\times[\underline{\sigma},\overline{\sigma}]}\!\lambda(\overline{\varsigma},\underline{\rho},\underline{b},a,\sigma)) & \beta\in[0,1),\,\underline{\rho}<0 \\
			(\argmax\limits_{\varsigma\in[\underline{\varsigma},\overline{\varsigma}]}\frac{1}{2}p(1-p)\beta^2\varsigma^2+p^2\beta(\beta-1)\underline{\rho}\frac{\underline{\sigma}}{\overline{a}}\varsigma,\overline{a},\underline{\sigma}) & \beta\in[0,1),\,\underline{\rho}>0 \\
			(\overline{\varsigma},\argmax\limits_{(a,\sigma)\in[\underline{a},\overline{a}]\times[\underline{\sigma},\overline{\sigma}]}\!\lambda(\overline{\varsigma},\overline{\rho},\underline{b},a,\sigma)) & \beta\in[\underline{\beta},0),\,\overline{\rho}>0 \\
			(\argmax\limits_{\varsigma\in[\underline{\varsigma},\overline{\varsigma}]}\frac{1}{2}p(1-p)\beta^2\varsigma^2+p^2\beta(\beta-1)\overline{\rho}\frac{\underline{\sigma}}{\overline{a}}\varsigma,\overline{a},\underline{\sigma}) & \beta\in[\underline{\beta},0),\,\overline{\rho}<0
		\end{cases}
	\end{equation}
	(Here, the symbol $'$ stands for the transpose of a vector).
	
\end{proposition}

For reasons similar to those discussed in Section \ref{sec:SV}, obtaining the explicit expressions of the optimal leverage ratio and the corresponding worst-case long-term growth rate under this model is difficult. Hence, we conduct numerical computations. We define
\begin{equation}
	\Lambda(\beta):=p\beta\mu^{*}(\beta)-\frac{1}{2}p(1-p)\beta^2\varsigma^*(\beta,\underline{\rho},\overline{\rho})^2-\lambda(\alpha^*(\beta,\underline{\rho},\overline{\rho})).
\end{equation}
In accordance with the discussion in Section \ref{sec:Heston}, a mesh size corresponding to a given error bound is obtained by the inequality given as
\begin{align}
	\left|\Lambda(\beta+h)-\Lambda(\beta)\right|\leq\left(p\,\overline{\mu}+p(1-p)\overline{\varsigma}^2\max\{|\underline{\beta}|,\overline{\beta}\}+\sup_{a,\sigma,\beta}|\lambda_{\beta}(a,\sigma,\beta)|\right)h.
\end{align}
\begin{figure}[h] 
	\centering\includegraphics[width=0.8\textwidth, height=0.4\textheight]{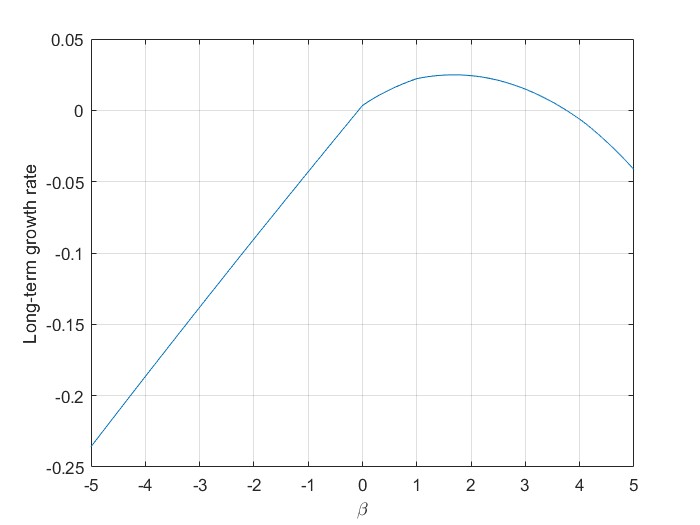}
	\caption{Long-term growth rate of the worst-case expected utility as a function of the leverage ratio $\beta$ under the Vasicek interest rate model.}\label{fig:Vasicek}
\end{figure}

We set $\beta\in[-5,5],\,p=0.5,\,[\underline{\mu},\overline{\mu}]=[0.06,0.1],\,[\underline{\varsigma},\overline{\varsigma}]=[0.08,0.25],\,[\underline{\rho},\overline{\rho}]=[-0.9,-0.5],\,[\underline{b},\overline{b}]=[0.06,0.1],\,[\underline{a},\overline{a}]=[6,9],\,[\underline{\sigma},\overline{\sigma}]=[0.2,0.5]$. Figure \ref{fig:Vasicek} illustrates the worst-case long-term growth rate of the expected utility as a function of $\beta$. The optimal leverage ratio and robust long-term growth rate are approximately 1.7 and 0.025, respectively, with error bounded to 0.01.

\begin{figure}[h]
	\begin{center} 
		\includegraphics[width=0.45\textwidth, height=0.225\textheight]{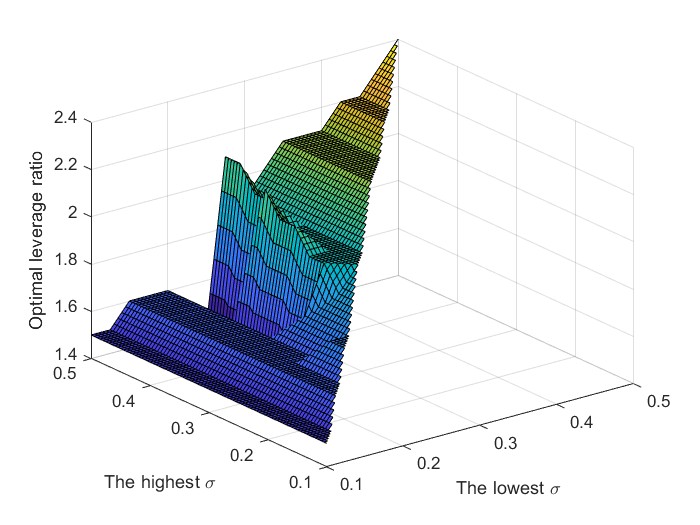} 
		\includegraphics[width=0.45\textwidth, height=0.225\textheight]{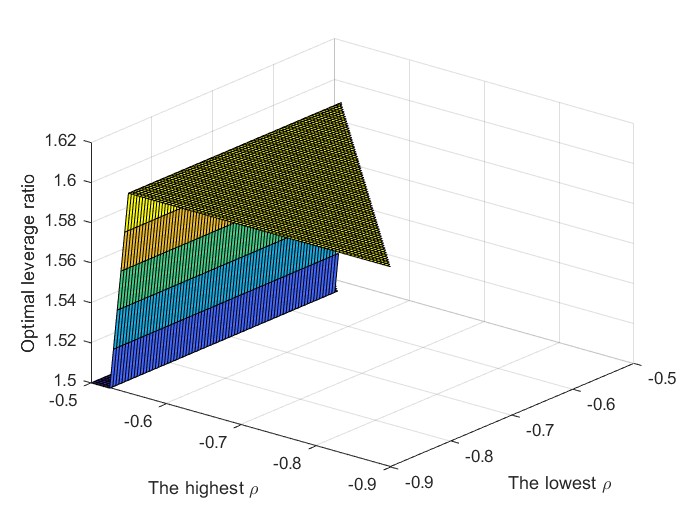}
		\caption{Corresponding optimal leverage ratios to different ranges of $\sigma$ and $\rho$.}\label{fig2:vasicek}
	\end{center}
\end{figure}

The optimal leverage ratios corresponding to various ranges of $\sigma$ and $\rho$ in the Vasicek model exhibit greater dynamism compared to those in the Heston model, as shown in Figure \ref{fig2:vasicek}. Specifically, the optimal leverage ratio is neither monotone in $\overline{\sigma}$ nor in $\underline{\sigma}$.

Furthermore, this stochastic interest rate model has extension similar to those in the stochastic volatility models.
\begin{cor}
	Let the reference process $X$ and the interest rate $r$ follow
	\begin{equation}
		\begin{aligned}
			&dX_t=\mu_tX_t\, dt+\varsigma_t\,X_t\,dW_t^{\rho(t)}\;,\\
			&dr_t=(b_0(t,r_t)-a\,r_t)\,dt+\sigma\,dB_t^{\rho(t)}\;,
		\end{aligned}
	\end{equation}
	where $\mu$ and $\varsigma$ are progressively measurable processes mapping to $[\underline{\mu},\overline{\mu}]$ and $[\underline{\varsigma},\overline{\varsigma}]$, respectively, and $b_0$ and $\rho$ range over $[\underline{b},\overline{b}]$ and $[\underline{\rho},\overline{\rho}]$, respectively. Then, Proposition \ref{prop:vasicek} holds for the LETF with the reference $X$ and the interest rate $r$.
\end{cor}

\subsection{Inverse GARCH interest rate}
\label{sec:r_I_GARCH}Assume that the reference price $X$ and short interest rate $r$ satisfy the SDEs 
\begin{equation}\label{eqn:inv_GARCH}
	\begin{aligned}
		&dX_t=\mu X_t\, dt+ \varsigma X_t\,dW_t\;,\\ 
		&dr_t=(b-a r_t)r_t\,dt+\sigma\,r_t\,dB_t\;, 
	\end{aligned} 
\end{equation}
where $W$ and $B$ are two  Brownian motions such that  $\langle W,B\rangle_t=\rho\,t$  with    $-1\leq\underline{\rho}\leq\rho\leq\overline{\rho}\leq 1.$ We assign uncertainties to $\alpha=(\mu,\varsigma,\rho,b,a,\sigma)\in[\underline{\alpha},\overline{\alpha}]=[\underline{\mu},\overline{\mu}]\times[\underline{\varsigma},\overline{\varsigma}]\times[\underline{\rho},\overline{\rho}]\times[\underline{b},\overline{b}]\times[\underline{a},\overline{a}]\times[\underline{\sigma},\overline{\sigma}]$ with $\underline{\mu},\underline{\varsigma},\underline{a},\underline{\sigma}>0$, and $\underline{b}-p|\beta|\overline{\varsigma}\,\overline{\sigma}>\overline{\sigma}^2/2$. Under the probability measure $\hat{\mbp}^{\alpha}$ defined by \eqref{RN:StocIR}, the process \eqref{eqn:inv_GARCH} follows
\begin{equation}
	\begin{aligned}
		dX_t&=(\mu+p\beta\varsigma^2)X_t\,dt+\varsigma X_t\,d\hat{W}_t,\\
		dr_t&=\left(b+p\beta\varsigma\rho\sigma-ar_t\right)r_tdt+\sigma\,d\hat{B}_t^{\alpha},\quad 0\leq t\leq T.
	\end{aligned}
\end{equation}
The comparison principle shows that
\begin{equation}
	\mathbb{E}^{\mbp^{\alpha}}[L_T^p]\geq e^{p\beta\mu^{*} T-p(1-p)\beta^2\varsigma^2T/2}\mathbb{E}^{\hat{\mbp}^{\mu^*,\varsigma,\rho,b^*,a^*,\sigma}}[e^{-p(\beta-1)\int_0^Tr_sds}],
\end{equation}
where
\begin{equation} \label{b,sigma:invGARCH}
	b^{*}=\begin{cases}
		\overline{b}, &  \beta\geq 1 \\ \underline{b}, & \beta<1
	\end{cases},\quad\mbox{ and }\quad
	a^{*}=\begin{cases}
		\underline{a}, &  \beta\geq 1 \\ \overline{a}, & \beta<1
	\end{cases}.
\end{equation}
Set $\alpha^*(\varsigma,\rho,\sigma)=(\mu^*,\varsigma,\rho,b^*,a^*,\sigma)$.

The eigenpair problem for the infinitesimal generator of $r$ is expressed as
$$-\lambda\phi(r)=\frac{1}{2}\sigma^2r^2\phi''(r)+(b^*+p\beta\varsigma\rho\sigma-a^*r)r\,\phi'(r)-p(\beta-1)r\,\phi(r),$$
and
\begin{equation*}
	\left(\lambda(\alpha^*(\varsigma,\rho,\sigma)),\phi(r)\right):=\left(-\frac{p(\beta-1)}{2a^*}\left(\frac{p(\beta-1)}{a^*}+1\right)\sigma^2+\frac{p^2\beta(\beta-1)\varsigma\rho}{a^*}\sigma+\frac{p(\beta-1)b^*}{a^*},r^{-\frac{p(\beta-1)}{a^*}} \right)
\end{equation*}
is one solution pair. By the martingale extraction method,
\begin{equation}
	V(T\,;\,\alpha^*(\varsigma,\rho,\sigma))=e^{-p(1-p)\beta^2\varsigma^2T/2-\lambda(\alpha^*(\varsigma,\rho,\sigma))T}r_0^{-p(\beta-1)/a^*}\mathbb{E}^{\hat{\mbq}^{\alpha^*\!(\sigma)}}[r_T^{p(\beta-1)/a^*}],
\end{equation}
where $\hat{\mbq}^{\alpha^*(\varsigma,\rho,\sigma)}$ is defined by 
\begin{equation} \label{RN:invGARCH}
	\frac{d\hat{\mathbb{Q}}^{\alpha^*(\varsigma,\rho,\sigma)}}{d\hat{\mathbb{P}}^{\alpha^*(\varsigma,\rho,\sigma)}}\biggl|_{\mathcal{F}_T}=\left(\frac{r_T}{r_0}\right)^{-\frac{p(\beta-1)}{a^*}}\!\exp{\left\{\lambda(\alpha^*(\varsigma,\rho,\sigma)) T-p(\beta-1)\int_0^T r_s\,ds\right\}}
\end{equation}
on $\mathcal{F}_T$ for each $T\geq 0$, and the process $r$ under $\hat{\mbq}^{\alpha^*(\varsigma,\rho,\sigma)}$ satisfies
$$dr_t=(b^{*}+p\beta\overline{\varsigma}\sigma-{p(\beta-1)\sigma^2/a^*} -a^*\,r_t)r_t\,dt+\sigma\,r_t\,dB_t^{\hat{\mbq}^{\alpha^*\!(\sigma)}}\;,$$
with a $\hat{\mbq}^{\alpha^*(\varsigma,\rho,\sigma)}$-Brownian motion
\begin{equation}
	B_t^{\hat{\mbq}^{\alpha^*(\varsigma,\rho,\sigma)}}=\frac{p(\beta-1)\sigma}{a^*}t+\hat{B}_t.
\end{equation}
The Radon-Nikodym derivative \eqref{RN:invGARCH} is well-defined using an argument similar to that used in Section \ref{sec:Uncertain_ref}.

We show that $\inf\limits_{\varsigma,\rho,\sigma,T}\mathbb{E}^{\hat{\mbq}^{\alpha^*\!(\sigma)}}[r_T^{p(\beta-1)/a^*}]>0$. Since $r^{-1}$ is a GARCH process expressed as a linear SDE, $r_t^{-1}$ can be written explicitly in the form (\citet{klebaner2012introduction})
\begin{equation}
	r_t^{-1}=r_0^{-1}e^{-(b^{*}+p\beta\varsigma\rho\sigma-{p(\beta-1)\sigma^2/a^*}-\sigma^2/2)t-\sigma B_t}+a^*\int_0^te^{-(b^{*}+p\beta\varsigma\rho\sigma-{p(\beta-1)\sigma^2/a^*}-\sigma^2/2)(t-s)-\sigma(B_t-B_s)}ds.
\end{equation}
\begin{mycase}
	\case $\beta\geq1$:
	Then $p(1-\beta)/a^*\leq 0$. By Jensen's inequality,
	\begin{align}
		\scriptstyle\mathbb{E}^{\hat{\mbq}^{\alpha^*\!(\sigma)}}[r_T^{-p(1-\beta)/a^*}]
		&\geq\,\scriptstyle\mathbb{E}^{\hat{\mbq}^{\alpha^*\!(\sigma)}}\![\scriptstyle r_0^{-1}e^{-(b^{*}+p\beta\varsigma\rho\sigma-\frac{p(\beta-1)\sigma^2}{a^*}-\frac{\sigma^2}{2})t-\sigma B_t}+a^*\!\int_0^te^{-(b^{*}+p\beta\varsigma\rho\sigma-\frac{p(\beta-1)\sigma^2}{a^*}-\frac{\sigma^2}{2})(t-s)-\sigma(B_t-B_s)}ds]^{\scriptstyle\frac{p(1-\beta)}{a^*}}\\&=\scriptstyle \left(r_0^{-1}e^{-(b^{*}+p\beta\varsigma\rho\sigma-\frac{p(\beta-1)\sigma^2}{a^*}-\sigma^2)t}+\frac{a^*}{b^{*}+p\beta\varsigma\rho\sigma-\frac{p(\beta-1)\sigma^2}{a^*}-\sigma^2}(1-e^{-(b^{*}+p\beta\varsigma\rho\sigma-\frac{p(\beta-1)\sigma^2}{a^*}-\sigma^2)t})\right)^{\frac{p(1-\beta)}{a^*}}.
	\end{align}
	
	\case $\beta<1$:
	Then, $p(1-\beta)/a^*>0$. Truncating the second term of $r_t^{-1}$ provides the trivial inequality given by
	\begin{align}
		\mathbb{E}^{\hat{\mbq}^{\alpha^*\!(\sigma)}}[r_T^{-p(1-\beta)/a^*}]&\geq\mathbb{E}^{\hat{\mbq}^{\alpha^*\!(\sigma)}}\! \left[r_0^{-p(1-\beta)/a^*}e^{-\frac{p(1-\beta)}{a^*}\left((b^{*}+p\beta\varsigma\rho\sigma-\frac{p(\beta-1)\sigma^2}{a^*}-\frac{\sigma^2}{2})t+\sigma B_t\right)}\right]\\
		&= r_0^{-p(1-\beta)/a^*}e^{-\frac{p(1-\beta)}{a^*}(b^{*}+p\beta\varsigma\rho\sigma-\frac{p(\beta-1)\sigma^2}{2a^*}-\sigma^2)t}.
	\end{align}
\end{mycase}

In both cases, the requirement holds. Additionally, if the condition `$b^{*}+p\beta\varsigma\rho\sigma-p(\beta-1)\sigma^2/a^*-\sigma^2>0$' holds, we can show by the argument discussed in Section \ref{sec:Uncertain_ref} that $\mathbb{E}^{\hat{\mbq}^{\alpha^*\!(\sigma)}}[r_T^{p(\beta-1)/a^*}]$ converges as $T\to\infty$. Thus,
\begin{equation}
	\lim_{T\to\infty}\frac{1}{T}\log\inf_{\alpha\in[\underline{\alpha},\overline{\alpha}]} V(T\,;\,\alpha)= \inf_{(\varsigma,\rho,\sigma)\in[\underline{\varsigma},\overline{\varsigma}]\times[\underline{\rho},\overline{\rho}]\times[\underline{\sigma},\overline{\sigma}]}\!-\frac{1}{2}p(1-p)\beta^2\varsigma^2-\lambda(\alpha^*(\varsigma,\rho,\sigma)),
\end{equation}
provided $\inf\limits_{(\varsigma,\rho,\sigma)\in[\underline{\varsigma},\overline{\varsigma}]\times[\underline{\rho},\overline{\rho}]\times[\underline{\sigma},\overline{\sigma}]}b^{*}+p\beta\varsigma\rho\sigma-\frac{p(\beta-1)\sigma^2}{a^*}-\sigma^2>0$.

Let $\alpha^*=(\mu^*,\varsigma^*,\rho^*,b^*,a^*,\sigma^*)$ achieve the infimum. Recall that $\mu^*,b^*,\sigma^*$ are already determined in \eqref{mu*} and \eqref{b,sigma:invGARCH}, respectively. \vspace{2mm}

\begin{mycase}
	\case $\beta\in[1,\overline{\beta}]$ and $\overline{\rho}>0$: $b^*=\overline{b},\,a^*=\underline{a},\,\varsigma^*=\overline{\varsigma},\,\rho^*=\overline{\rho}$, and $$ \sigma^*\in\argmin_{\sigma\in[\underline{\sigma},\overline{\sigma}]}\frac{p(\beta-1)}{2\underline{a}}\left(\frac{p(\beta-1)}{\underline{a}}+1\right)\sigma^2-\frac{p^2\beta(\beta-1)\overline{\varsigma}\overline{\rho}}{\underline{a}}\sigma.$$
	
	\case $\beta\in[1,\overline{\beta}]$ and $\overline{\rho}<0$: $b^*=\overline{b},\,a^*=\underline{a},\,\rho^*=\overline{\rho},\,\sigma^*=\underline{\sigma}$, and $$ \varsigma^*\in\argmin_{\varsigma\in[\underline{\varsigma},\overline{\varsigma}]}-\frac{1}{2}p(1-p)\beta^2\varsigma^2-\frac{p^2\beta(\beta-1)\overline{\rho}\underline{\sigma}}{\underline{a}}\varsigma.$$

	\case $\beta\in[0,1)$ and $\underline{\rho}<0$: $b^*=\underline{b},\,a^*=\overline{a},\,\varsigma^*=\overline{\varsigma},\,\rho^*=\underline{\rho}$, and $$ \sigma^*\in\argmin_{\sigma\in[\underline{\sigma},\overline{\sigma}]}\frac{p(\beta-1)}{2\overline{a}}\left(\frac{p(\beta-1)}{\overline{a}}+1\right)\sigma^2-\frac{p^2\beta(\beta-1)\overline{\varsigma}\underline{\rho}}{\overline{a}}\sigma.$$
	
	\case $\beta\in[0,1)$ and $\underline{\rho}>0$: $b^*=\underline{b},\,a^*=\overline{a},\,\rho^*=\underline{\rho}$, and
	\begin{equation}
		(\varsigma^*,\sigma^*)=\argmin_{(\varsigma,\sigma)\in[\underline{\varsigma},\overline{\varsigma}]\times[\underline{\sigma},\overline{\sigma}]}\!-\frac{1}{2}p(1-p)\beta^2\varsigma^2+\frac{p(\beta-1)}{2\overline{a}}\left(\frac{p(\beta-1)}{\overline{a}}+1\right)\sigma^2-\frac{p^2\beta(\beta-1)\underline{\rho}}{\overline{a}}\varsigma\sigma.
	\end{equation}
	
	\case $\beta\in[\underline{\beta},0)$ and $\overline{\rho}>0$: $b^*=\underline{b},\,a^*=\overline{a},\,\varsigma^*=\overline{\varsigma},\,\rho^*=\overline{\rho}$, and $$ \sigma^*\in\argmin_{\sigma\in[\underline{\sigma},\overline{\sigma}]}\frac{p(\beta-1)}{2\overline{a}}\left(\frac{p(\beta-1)}{\overline{a}}+1\right)\sigma^2-\frac{p^2\beta(\beta-1)\overline{\varsigma}\overline{\rho}}{\overline{a}}\sigma.$$
	
	\case $\beta\in[\underline{\beta},0)$ and $\overline{\rho}<0$:  $b^*=\underline{b},\,a^*=\overline{a},\,\rho^*=\overline{\rho}$, and
	\begin{equation}
		(\varsigma^*,\sigma^*)=\argmin_{(\varsigma,\sigma)\in[\underline{\varsigma},\overline{\varsigma}]\times[\underline{\sigma},\overline{\sigma}]}\!-\frac{1}{2}p(1-p)\beta^2\varsigma^2+\frac{p(\beta-1)}{2\overline{a}}\left(\frac{p(\beta-1)}{\overline{a}}+1\right)\sigma^2-\frac{p^2\beta(\beta-1)\overline{\rho}}{\overline{a}}\varsigma\sigma.
	\end{equation}
	
\end{mycase}

\begin{proposition} \label{prop:inv_GARCH}
	Let $0<\underline{\mu}\leq\overline{\mu},\,0<\underline{\varsigma}\leq\overline{\varsigma},\,0<\underline{a}\leq\overline{a},\,0<\underline{\sigma}\leq\overline{\sigma},\,\overline{\sigma}^2/2+p|\beta|\overline{\varsigma}\,\overline{\sigma}<\underline{b}\leq\overline{b}$, and $(X^{\alpha},r^{\alpha})$ be the process \eqref{eqn:inv_GARCH} with set of parameters $\alpha=(\mu,\varsigma,\rho,b,a,\sigma)$ ranging over $[\underline{\alpha},\overline{\alpha}]=[\underline{\mu},\overline{\mu}]\times[\underline{\varsigma},\overline{\varsigma}]\times[\underline{\rho},\overline{\rho}]\times[\underline{b},\overline{b}]\times[\underline{a},\overline{a}]\times[\underline{\sigma},\overline{\sigma}]$.
	Then, the long-term growth rate of the worst-case expected utility of the LETF $L^{\alpha}=(L_t^{\alpha})_{t\geq 0}$ with reference process and interest rate $X^{\alpha}$ and $r^{\alpha}$ respectively, is given by
	\begin{align}
		\lim_{T\to\infty}\frac{1}{T}\log{\inf_{\alpha\in[\underline{\alpha},\overline{\alpha}]}\mathbb{E}^{\mbp^{\alpha}}[L_T^p]}&=p\beta\mu^{*}(\beta)-\frac{1}{2}p(1-p)\beta^2\varsigma^*(\beta,\underline{\rho},\overline{\rho})^2-\frac{p(\beta-1)b^*(\beta)}{a^*(\beta)}\\&\,\,+\frac{p(\beta-1)}{2a^*(\beta)}\left(\frac{p(\beta-1)}{a^*(\beta)}+1\right)\sigma^*(\beta,\underline{\rho},\overline{\rho})^2-\frac{p^2\beta(\beta-1)}{a^*(\beta)}(\varsigma^*\sigma^*)(\beta,\underline{\rho},\overline{\rho}),
	\end{align}
	where
	\begin{equation}
		\mu^{*}(\beta)=\begin{cases}\underline{\mu}, &  \beta\geq 0 \\ \overline{\mu}, & \beta<0
		\end{cases},\,\rho^*(\beta)=\begin{cases}
			\overline{\rho}, & \beta\in[\underline{\beta},0)\cup[1,\overline{\beta}] \\ \underline{\rho}, & \beta\in[0,1) 
		\end{cases},\, (b^*(\beta),a^*(\beta))=\begin{cases}(\overline{b},\underline{a}), &  \beta\geq 1 \\ (\underline{b},\overline{a}), & \beta<1\end{cases},
	\end{equation}
	and
	\begin{equation}
		\begin{pmatrix}
			\varsigma^*(\beta,\underline{\rho},\overline{\rho})\\ \sigma^*(\beta,\underline{\rho},\overline{\rho})
		\end{pmatrix}'
		=\begin{cases}
			(\overline{\varsigma},\argmin\limits_{\sigma\in[\underline{\sigma},\overline{\sigma}]}\frac{p(\beta-1)}{2\underline{a}}\left(\frac{p(\beta-1)}{\underline{a}}+1\right)\sigma^2-\frac{p^2\beta(\beta-1)\overline{\varsigma}\overline{\rho}}{\underline{a}}\sigma) & \beta\in[1,\overline{\beta}],\,\overline{\rho}>0 \\
			(\argmin\limits_{\varsigma\in[\underline{\varsigma},\overline{\varsigma}]}-\frac{1}{2}p(1-p)\beta^2\varsigma^2-\frac{p^2\beta(\beta-1)\overline{\rho}\underline{\sigma}}{\underline{a}}\varsigma,\underline{\sigma}) & \beta\in[1,\overline{\beta}],\,\overline{\rho}<0 \\ 
			(\overline{\varsigma},\argmin\limits_{\sigma\in[\underline{\sigma},\overline{\sigma}]}\frac{p(\beta-1)}{2\overline{a}}\left(\frac{p(\beta-1)}{\overline{a}}+1\right)\sigma^2-\frac{p^2\beta(\beta-1)\overline{\varsigma}\underline{\rho}}{\overline{a}}\sigma) & \beta\in[0,1),\,\underline{\rho}<0 \\
			(\argmax\limits_{(\varsigma,\sigma)\in[\underline{\varsigma},\overline{\varsigma}]\times[\underline{\sigma},\overline{\sigma}]}\!\frac{1}{2}(1-p)\beta^2\varsigma^2-\frac{\beta-1}{2\overline{a}}\!(\frac{p(\beta-1)}{\overline{a}}+1)\!\sigma^2+\frac{p\beta(\beta-1)\underline{\rho}}{\overline{a}}\varsigma\sigma) & \beta\in[0,1),\,\underline{\rho}>0 \\
			(\overline{\varsigma},\argmin\limits_{\sigma\in[\underline{\sigma},\overline{\sigma}]}\frac{p(\beta-1)}{2\overline{a}}\left(\frac{p(\beta-1)}{\overline{a}}+1\right)\sigma^2-\frac{p^2\beta(\beta-1)\overline{\varsigma}\overline{\rho}}{\overline{a}}\sigma) & \beta\in[\underline{\beta},0),\,\overline{\rho}>0 \\
			(\argmax\limits_{(\varsigma,\sigma)\in[\underline{\varsigma},\overline{\varsigma}]\times[\underline{\sigma},\overline{\sigma}]}\!\frac{1}{2}(1-p)\beta^2\varsigma^2-\frac{\beta-1}{2\overline{a}}\!(\frac{p(\beta-1)}{\overline{a}}+1)\!\sigma^2+\frac{p\beta(\beta-1)\overline{\rho}}{\overline{a}}\varsigma\sigma) & \beta\in[\underline{\beta},0),\,\overline{\rho}<0
		\end{cases}
	\end{equation}
	(Here, the symbol $'$ stands for the transpose of a vector), provided that \[\inf\limits_{(\varsigma,\rho,\sigma)\in[\underline{\varsigma},\overline{\varsigma}]\times[\underline{\rho},\overline{\rho}]\times[\underline{\sigma},\overline{\sigma}]}b^{*}(\beta)+p\beta\varsigma\rho\sigma-\frac{p(\beta-1)\sigma^2}{a^*(\beta)}-\sigma^2>0\] for all $\beta\in[\underline{\beta},\overline{\beta}]$.
\end{proposition}

The problem associated with finding the optimal leverage ratio can also be solved numerically as aforementioned.
\begin{cor}
	Let the reference process $X$ and interest rate $r$ follow
	\begin{equation}
		\begin{aligned}
			&dX_t=\mu_tX_t\, dt+\varsigma_t\,X_t\,dW_t^{\rho(t)}\;,\\
			&dr_t=(b_0(t,r_t)-a_0(t,r_t)\,r_t)\,dt+\sigma\,dB_t^{\rho(t)}\;,
		\end{aligned}
	\end{equation}
	where $\mu$ and $\varsigma$ are progressively measurable processes mapping to $[\underline{\mu},\overline{\mu}]$ and $[\underline{\varsigma},\overline{\varsigma}]$, respectively, and $b_0,\,a_0$, and $\rho$ take values in $[\underline{b},\overline{b}],\,[\underline{a},\overline{a}]$, and $[\underline{\rho},\overline{\rho}]$, respectively. Then, Proposition \ref{prop:inv_GARCH} holds for the LETF with the reference $X$ and interest rate $r$.
\end{cor}

\section{Conclusions}
\label{sec:conclusion}
Expanding the analysis in \citet{Leung2017} on the long-term growth rate of expected utility of LETF, we have conducted an analysis of the worst-case scenario for an agent holding an LETF in this study. Along with the previously introduced martingale extraction method, the comparison principle is employed to determine the worst-case scenario. Various models, including stochastic volatility and interest rate models are covered. The optimal leverage ratio varies across models and depends on parameter relationships. Particularly, the numerical experiment illustrates a strong dependency of the optimal leverage ratio on the correlation and volatility ranges of the volatility process under the Heston model. Additionally, investing in the LETF is observed to be unfavorable for agents unless the worst expected rate of return of the reference asset is significantly higher than the interest rate, as expected.

This paper provides ETF issuers, portfolio managers, and regulators with a useful and flexible framework to understand the robust growth rate of LETFs in the long run. The approach discussed herein can also be applied to reference indexes or assets with different dynamics. Therefore, the LETFs can be from asset classes other than equities, including interest rates, commodities, and currencies.

Future research directions include considering alternative stochastic models for the reference asset, such as jump-diffusion and regime-switching models. The combined effects of leverage and asset dynamics on the long-run growth rate of LETFs could provide useful insights for fund selection and risk management.

\section*{Acknowledgments.}
Hyungbin Park was supported by the National Research Foundation of Korea (NRF) grant funded by the Korea government (MSIT) (No. 2021R1C1C1011675 and No. 2022R1A5A6000840). Financial support from the Institute for Research in Finance and Economics of Seoul National University is gratefully acknowledged.

\appendices

\begin{small}
\begin{spacing}{0.5}
\bibliographystyle{abbrvnat}
\bibliography{LETF2022}
\end{spacing}
\end{small}

\end{document}